\documentclass[pra,showpacs,groupedaddress,amssymb,twocolumn,notitlepage,nofootinbib,floatfix,superscriptaddress]{revtex4-1}
\usepackage{graphicx,amsmath,dsfont,physics}
\usepackage{amsthm}
\usepackage[T1]{fontenc}
\usepackage[dvipsnames]{xcolor}
\usepackage{subcaption}
\usepackage[english]{babel}
\usepackage{animate}
\graphicspath{{Figures2/}}
\usepackage{tabularx}
\usepackage{enumerate,mathtools}
\usepackage{comment}
\usepackage{hyperref}
\hypersetup{
    colorlinks=true,
    linkcolor=red,
    citecolor=blue,
    urlcolor=cyan,
    pdfauthor={Teresa CAI}
}

\usepackage[most]{tcolorbox}
\usepackage[capitalize]{cleveref}
\usepackage{tikz}
\usetikzlibrary{arrows.meta, calc, decorations.pathmorphing, shapes, positioning,shapes.geometric} 
\usepackage{braket}

\newcommand{\drawDetector}{
  \pgfpathellipse{\pgfpoint{0cm}{0cm}}{\pgfpoint{0.25 cm/4}{0}}{\pgfpoint{0}{0.25 cm/4*sin(15)}}
  \pgfusepath{stroke,fill}  

  \color{black}
  \pgfpathmoveto{\pgfpoint{0.25cm}{0cm}}
  \pgfpatharc{0}{180}{0.25cm}
  \pgfpatharc{180}{0}{0.25 cm and {0.25 cm*sin(15)}}
  \pgfusepath{stroke, fill}

  \pgfpathellipse{\pgfpoint{0cm}{0cm}}{\pgfpoint{0.25 cm}{0}}{\pgfpoint{0}{0.25 cm*sin(15)}}
  \pgfusepath{stroke}
}
\pgfdeclareshape{BSM_shape}{
  \anchor{center}{\pgfpoint{0cm}{0cm}}
  \anchor{north}{\pgfpoint{0cm}{sin(45) * 0.5 cm}}
  \anchor{north west}{\pgfpoint{-cos(45) * 0.8 cm}{sin(45) * 0.75 cm + 0.25 cm}}
  \anchor{north east}{\pgfpoint{cos(45) * 0.8 cm}{sin(45) * 0.75 cm + 0.25 cm}}
  \anchor{south}{\pgfpoint{0cm}{-sin(45) * 0.5 cm}}
  \anchor{south west}{\pgfpoint{-cos(45) * 0.25 cm}{-sin(45) * 0.25 cm}}
  \anchor{south east}{\pgfpoint{cos(45) * 0.25 cm}{-sin(45) * 0.25 cm}}
  
  \backgroundpath{
    \pgfscope
      \pgftransformshift{\pgfpoint{-cos(45) * 0.5 cm}{sin(45) * 0.5 cm}}
      \pgftransformrotate{45}
      \drawDetector
    \endpgfscope
    \pgfscope
      \pgftransformshift{\pgfpoint{cos(45) * 0.5 cm}{sin(45) * 0.5 cm}}
      \pgftransformrotate{-45}
      \drawDetector
    \endpgfscope

    \pgfpathmoveto{\pgfpoint{-cos(45) * 0.5 cm}{sin(45) * 0.5 cm}}
    \pgflineto{\pgfpoint{cos(45) * 0.25 cm}{-sin(45) * 0.25 cm}}
    \pgfpathmoveto{\pgfpoint{cos(45) * 0.5 cm}{sin(45) * 0.5 cm}}
    \pgflineto{\pgfpoint{-cos(45) * 0.25 cm}{-sin(45) * 0.25 cm}}
    \pgfusepath{stroke}

    \pgfsetstrokecolor{black}
    \pgfsetfillcolor{cyan!70!white}
    \pgfsetfillopacity{0.3}
    \pgfpathmoveto{\pgfpoint{0 cm}{sin(45)* 0.5 cm}}
    \pgflineto{\pgfpoint{cos(45)* 0.5 cm}{0 cm}}
    \pgflineto{\pgfpoint{0 cm}{-sin(45)* 0.5 cm}}
    \pgflineto{\pgfpoint{-cos(45)* 0.5 cm}{0 cm}}
    \pgfpathclose
    \pgflineto{\pgfpoint{0 cm}{-sin(45)* 0.5 cm}} 
    \pgfusepath{fill, stroke}
  }
}
\pgfkeys{
  /pgf/decoration/loops/radius A/.initial=0.25cm,
  /pgf/decoration/loops/radius B/.initial=0.2cm,
  /pgf/decoration/loops/extra loops/.initial=0,
  /pgf/decoration/loops/side/.initial=1,
}
\pgfdeclaredecoration{loop shape}{initial}{
  \state{initial}[width=0pt,next state=final]{
        \pgfmathsetlength{\pgf@xa}{\pgfkeysvalueof{/pgf/decoration/loops/radius A}}
        \pgfmathsetlength{\pgf@xb}{\pgfkeysvalueof{/pgf/decoration/loops/radius B}}
        \pgfmathsetmacro{\side}{\pgfkeysvalueof{/pgf/decoration/loops/side}}
        \pgfpathlineto{\pgfpoint{\pgfdecoratedpathlength-\pgf@xa}{0}}
        \pgfpatharc{-\side*90}{0}{\pgf@xa}
  }
  \state{final}{
        \pgfmathsetlength{\pgf@xa}{\pgfkeysvalueof{/pgf/decoration/loops/radius A}}
        \pgfmathsetlength{\pgf@xb}{\pgfkeysvalueof{/pgf/decoration/loops/radius B}}
        \pgfmathsetmacro{\side}{\pgfkeysvalueof{/pgf/decoration/loops/side}}  
        \pgfmathsetmacro{\numl}{\pgfkeysvalueof{/pgf/decoration/loops/extra loops}}
        \ifnum\numl>0
        \foreach \x in {1,...,\numl}{
            \pgfpatharc{0}{\side*360}{\pgf@xb}
            \pgflineto{\pgfpointadd{\pgfpoint{\pgfdecoratedpathlength}{\side*\pgf@xa}}{\x*\side*\pgfpoint{0}{0.025 cm}}}    
        }
        \fi
        \pgfpatharc{0}{\side*180}{\pgf@xb}
        \pgflineto{\pgfpoint{\pgfdecoratedpathlength-2*\pgf@xb}{\side*\pgf@xa}}
        \pgfpatharc{180}{180+\side*90}{\pgf@xa}
        \pgfpathlineto{\pgfpointdecoratedpathlast}
  }
}

\newcounter{protocol}

\usepackage{color}
\usepackage{amsthm}
\usepackage{amsmath,amssymb,amsbsy}

\usepackage{xcolor}
\pagecolor{white}
\usepackage[english]{babel}

\usepackage{calrsfs}

\newcommand{\lettersection}[1]{\emph{#1.---}}

\newcommand{\eea}{\end{eqnarray}}
\newcommand{\bea}{\begin{eqnarray}}
\newcommand{\ee}{\end{equation}}
\newcommand{\be}{\begin{equation}}

\begin{document}
\title{High-Fidelity Remote Graph State Preparation for Blind Quantum Computation}

\author{Jiawei Cai}
\affiliation{Leiden Institute of Physics\\ Leiden University, 2333 CA Leiden, The Netherlands}

\author{Rex Fleur}
\affiliation{MasonQ, Herengracht 566, 1017 CH Amsterdam, The Netherlands}

\author{Benedikt Tissot}
\affiliation{Center for Hybrid Quantum Networks (Hy-Q), Niels Bohr Institute,\\ University of Copenhagen, Jagtvej 155A, Copenhagen DK-2200, Denmark}

\author{Wolfgang Löffler}
\affiliation{Leiden Institute of Physics\\ Leiden University, 2333 CA Leiden, The Netherlands}

\author{Tzula B. Propp}
\affiliation{Leiden Institute of Physics\\ Leiden University, 2333 CA Leiden, The Netherlands}
\affiliation{MasonQ, Herengracht 566, 1017 CH Amsterdam, The Netherlands}

\begin{abstract}
Measurement-based quantum computation (MBQC) relies on entangled graph states, yet existing remote state preparation (RSP) protocols prepare only separable states, requiring subsequent entangling gates on the remote server. Here, we introduce Remote Graph State Preparation (RGSP), a framework that prepares arbitrary graph states directly from a single high-dimensional photonic qudit. By encoding multiple qubits and their graph connectivity into the photon's structured phase profile, RGSP can reduce or, for small computations, completely eliminate the need for server-side entangling operations among qubits. We show that under a reasonable noise model the resulting state fidelity is independent of the graph topology. Under cumulative fiber phase drift, we demonstrate that this fidelity is significantly enhanced by a ``highest-weight-first'' mode reordering strategy. Finally, we show that RGSP enables reduced-SWAP universal blind quantum computing on arbitrary topologies, reducing qubit overhead e.g., for a six-qubit Quantum Fourier Transform from $420$ RSP qubits with a standard brickwork topology to $48$ RSP qubits. These results establish RGSP as a resource-efficient, topology-invariant primitive for quantum-secured cloud computing.
\end{abstract}

\maketitle
\vspace{-1em}
\lettersection{Introduction}
\label{intro}
Large-scale fault-tolerant quantum computers are expected to be accessed primarily through the cloud, as building and maintaining quantum hardware remains technologically demanding and prohibitively expensive~\cite{Sarma2022, Alcazar2020, Proctor2022, Amaro2022, Kirsopp2022}. In this delegated computing model, a client delegates a quantum computation to a remote quantum server and, if they possess the ability to remotely prepare qubit states on the server, may keep their input, algorithm, and output private i.e., blind quantum computation (BQC)~\cite{broadbentUniversalBlindQuantum2009a,kashefiMultipartyDelegatedQuantum2017}. Measurement-based quantum computation (MBQC) is particularly suited to implementing BQC because once clients prepare randomly rotated
single-qubit states and send classical measurement
instructions, while the server entangles and measures the universal graph-state resource~\cite{Briegel2001, Raussendorf2003}. 
Efficient remote preparation of many qubit product and graph states is therefore a fundamental challenge for secure quantum networks.

To address this, remote state preparation (RSP) protocols have been developed to quickly and reliably generate target states on a remote quantum processor using quantum links and classical communication~\cite{bennett2001remote,leung2003oblivious}. Many RSP implementations prepare individual
qubits~\cite{drmota_verifiable_2024,Babichev2004,damSingleclickProtocolsRemote2025,Peters2005,Rosenfeld2007,Iuliano2024}: in the absence of quantum error correction, sequential register preparation exposes early qubits
to memory decoherence while the remaining qubits are
prepared~\cite{Gorin2007,Brennen2008,Baumer2024}. High-dimensional photonic
encoding~\cite{BechmannPasquinucci2000,bechmannpasquinucci2000a}
addresses this waiting-time overhead by encoding $n$
qubits across $2^n$ coherent optical modes of one photon and
heralding their preparation
jointly~\cite{propp2026remotely}. Reflection-based entanglement generation has also
been investigated
previously~\cite{nemoto-2014-photon_archit_scalab_quant_infor_proces_diamon,Omlor2025,Zheng2022}. 

In this Letter, we develop a protocol for \emph{Remote Graph State Preparation (RGSP)}, allowing the preparation of arbitrary graphs on a remote quantum cloud using a single high-dimensional photon\footnote{Extensions using weak coherent
pulses~\cite{Dunjko2012,Jiang2019} are also possible. For brevity, we focus here only on the single-photon implementations.}. Extending product-state preparation
$\bigotimes_{i=1}^{n}|+_{\theta_i}\rangle$ to graph states
provides the entangled resources required for graph-based
universal MBQC~\cite{rimock_generalized_2026,drmota_verifiable_2024}.
Within a blind protocol, this allows client-programmed
connectivity without disclosing the graph topology,
rather than requiring a fixed brickwork or cluster
resource~\cite{Briegel2001,Nielsen2006,Morimae2012,Takeuchi2019}. Directly preparing the required connectivity can reduce the SWAP routing overhead associated with nonlocal interactions on nearest-neighbour
architectures~\cite{Ma2024}, thereby reducing graph
size, circuit depth, and opportunities for error.

RGSP also reduces (or, for small computations, eliminates) the need for direct entangling gates on the server side, shifting the computational complexity from server-side graph-state preparation to client-side photonic engineering. We show that the preparation fidelity is independent of graph topology, depends only on optical phase noise, and naturally enables highly efficient, reduced-SWAP universal blind quantum computation on arbitrary topologies. Beyond BQC, we also discuss quantum
position verification~\cite{escolàfarràs2025quantumpositionverificationshot,bartusek2026trackqubitsspacetime}
and multi-party delegated
computation~\cite{polacchiMulticlientDistributedBlind2023},
highlighting RGSP as a general resource-generation primitive for
remote quantum computation.

\begin{figure}
    \centering
    \includegraphics[width=\columnwidth]{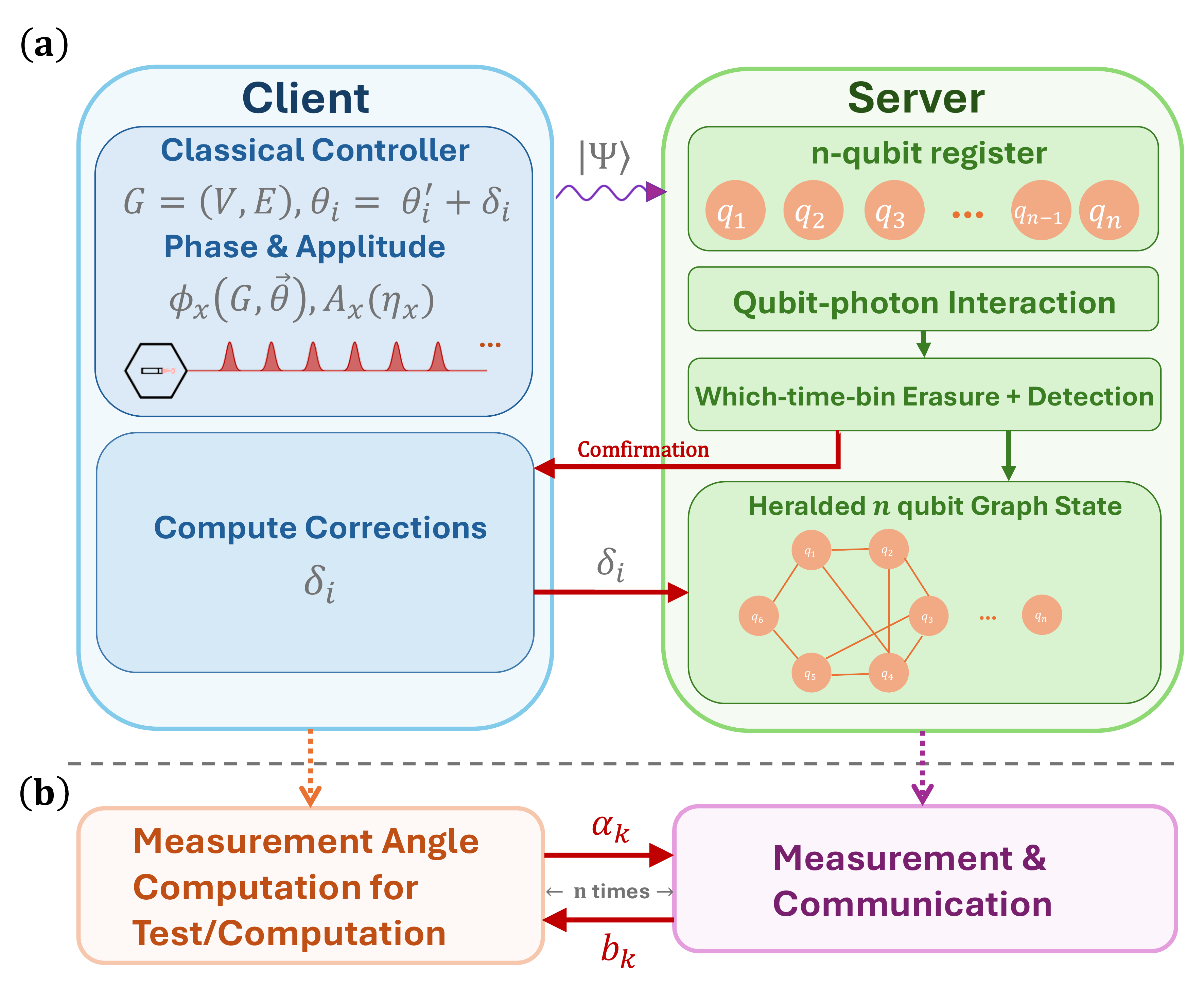}
    \caption{Overview of Remote Graph State Preparation (RGSP) for measurement-based blind quantum computation. (a) The client encodes both the local phases and the graph connectivity into the phase function of a single high-dimensional photon. Photon--matter interactions and which-mode erasureherald the remote graph state, up to known local
    corrections. (b) Adaptive single-qubit measurements consume the prepared resource for MBQC.
    See Appendix~\ref{app:RGSP_protocol}.}
    \label{fig:overview-grsp-bqc}
\end{figure}

\lettersection{Theory}
\label{theory} We consider single photon distributed coherently over $d=2^n$
temporal modes encodes an $n$-qubit basis through
$\vec{x}\in\{0,1\}^n$~\cite{propp2026remotely}:
\begin{equation}
|\psi_c\rangle=\sum_{\vec{x}\in\{0,1\}^n}c_xe^{i\phi_x}|1\rangle_x,
\label{eq:photon-state-clean}
\end{equation}
where $|1\rangle_x=a_x^\dagger|\mathrm{vac}\rangle$ and
$\sum_x|c_x|^2=1$. The amplitudes can compensate for
mode-dependent loss (Appendix~\ref{app:noise}). For a target graph state $|G\rangle$ with adjacency matrix $A$ and local
rotation angles $\vec{\theta}$, the client programs:
\begin{equation}
\phi_x
=
\sum_i\theta_i x_i
+
\pi\sum_{i<j}A_{ij}x_i x_j
\quad(\mathrm{mod}\;2\pi).
\label{eq:phase-rule}
\end{equation}

The linear and quadratic terms specify local rotations
and graph connectivity, respectively. As this qudit propagates through the server's quantum register (Fig.~\ref{fig:overview-grsp-bqc}), the phase pattern is transferred to the matter qubits via a fixed light-matter interaction sequence \cite{propp2026remotely}. Erasing the which-time-bin information via a quantum Fourier transform (QFT) or a time lens subsequently projects the atomic register into the target entangled state $|G\rangle$ without requiring physical two-qubit gates on the server side. Consequently, changing $A$ therefore selects arbitrary graph connectivity within the same interaction sequence
(Fig.~\ref{fig:arbitrary_graphs}); see
Appendices~\ref{app: derivation_of_RGSP}
and~\ref{app:RGSP_protocol}.


\begin{figure}[t]
    \centering
    \includegraphics[width=0.7\columnwidth]{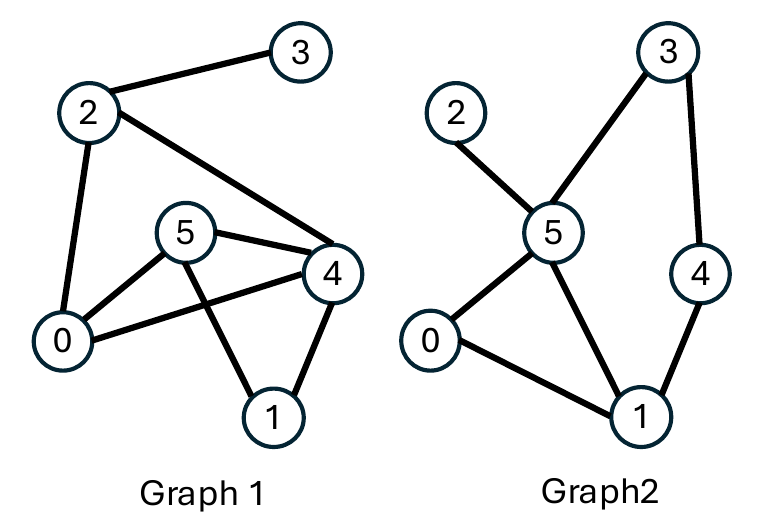}
    \caption{Representative graph states illustrating that RGSP directly prepares arbitrary target graph states by changing only the adjacency matrix $A_{ij}$ in Eq.~(\ref{eq:phase-rule}).}
    \label{fig:arbitrary_graphs}
\end{figure}

\paragraph*{Topology-independent fidelity.}
\label{Para:topology}
An immediate consequence in Eq.~(\ref{eq:phase-rule}) is that the preparation fidelity is independent of the encoded graph topology.

For an ideal normalized state with basis weights $w_x$ under fiber phase noise, stochastic phase errors $\epsilon_x$ give:
\begin{equation}
F
=
\left|\sum_x w_x e^{i\epsilon_x}\right|^2,
\qquad
w_x=|c_x|^2.
\label{eq:fidelity}
\end{equation}
The programmed phases $\phi_x$ cancel in the
ideal--noisy overlap $\langle \psi_{\text{ideal}} | \psi_{\text{noisy}} \rangle$ (Appendix~\ref{app:fidelity}). Under identical experimental conditions, a highly connected graph (such as a QFT resource state) can be prepared with the exact same fidelity as a simple one-dimensional chain, and without revealing to the server any information about the topology of the graph prepared except for its size. This topological invariance shifts the scalability limit of secure delegated quantum computing from the complexity of the target algorithm to the phase stability of the optical link. 

Furthermore, because the fidelity in Eq.~\eqref{eq:fidelity} is invariant under
$\epsilon_x\mapsto\epsilon_x+\epsilon_0$, so common-mode
phase drift does not affect this fidelity. As in dual-rail photonic encoding~\cite{kokLinearOpticalQuantum2007},
the relevant quantity is relative phase.
RGSP therefore requires relative phase stability between the temporal modes of a single photon, over the relevant pulse-train duration, rather than stabilization of a
shared absolute optical phase between distant nodes
as required in interferometric single-click
schemes~\cite{damSingleclickProtocolsRemote2025,Tissot2026PhaseSync}.

\paragraph*{Noise models.}
\label{para:noise}
Having established that the fidelity is independent of the encoded graph topology, the remaining limitation of RGSP arises from phase noise during transmission. We consider two experimentally relevant fiber phase-noise models that capture different stabilization regimes.
\begin{enumerate}
    \item \emph{Independent and identically distributed (IID) phase noise}: This model assumes independent Gaussian phase fluctuations $\epsilon_x \sim \mathcal{N}(0, \sigma^2)$ between temporal modes. This approximation is appropriate when active phase stabilization suppresses long-term drift, which describes uncorrelated residual errors.
    \item \emph{Cumulative phase noise}: This model describes a discretized Wiener process (Brownian motion), where cumulative drift with independent Gaussian increments between successive temporal modes, capturing the gradual phase evolution expected in passively stabilized optical fibers.
\end{enumerate}

Because the mode-dependent transmission losses scale exponentially with the Hamming weight $H(\vec{x})$, compensation requires assigning exponentially larger weights $w_x$ to high-Hamming-weight modes to equalize their conditional contributions at the server register (see Fig.~\ref{fig:loss_compensated_weights} in the Supplemental Material). Under cumulative Wiener phase noise, the phase variance of the $k$-th temporal mode scales as $\langle \delta\phi_k^2 \rangle = k \sigma_{\phi}^2$, meaning later modes suffer from larger phase drift. Since the overall fidelity in Eq.~\eqref{eq:fidelity} is dominated by a weighted sum, any dephasing in the heavily weighted modes rapidly degrades the fidelity.

\begin{figure}
    \centering
    \hspace*{-0.65cm}    
    \includegraphics[width=1\columnwidth]{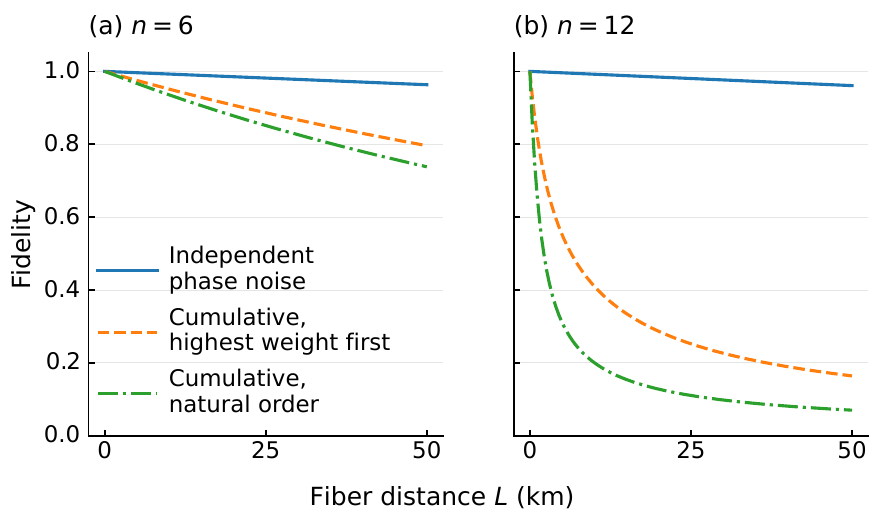}
    \caption{Output graph-state fidelity as a function of fiber transmission distance for $n=6$ (left) and $n=12$ (right).
    Blue curves assume independent Gaussian phase fluctuations between temporal modes, while green curves correspond to cumulative phase drift modeled as a Wiener process. Orange curves show the cumulative-noise model with the proposed highest-weight-first pulse ordering.
    The model uses
    $\sigma_\phi^2(L)=2D_\phi L$, with
    $D_\phi=4\times10^{-4}\,\mathrm{rad}^2/\mathrm{km}$.
    The interpretation of the ordering gain depends
    on the mode weights used in
    Eq.~\eqref{eq:fidelity}.
    }
    \label{fig:fidelity_distance}
\end{figure}

To mitigate this, we propose a \emph{highest-weight-first (High-H First)} mode-ordering strategy. By reordering the pulse sequence to transmit high-Hamming-weight modes first, we ensure that the most critical, heavily weighted components are transmitted at the beginning of the pulse train, before the Wiener process accumulates significant drift. As showed in Fig.~\ref{fig:fidelity_distance}, this simple mode-ordering strategy improves the prepared graph-state fidelity without modifying either the optical hardware or the RGSP protocol, with the performance gain becoming increasingly pronounced for larger system sizes.

\paragraph*{Graph-to-graph transformations.}
\label{para:g2g}
The phase-function representation also provides a unified description of transformations between graph states e.g., by an intermediate client that can modify the phase of each pulse. Let $\Delta$ be a symmetric binary matrix with zero diagonal, specifying an edge-toggle pattern: $\Delta_{ij}=1$ removes edge $(i,j)$ if present and adds it if absent. Applying the quadratic
phase update:
\begin{equation}
\phi_{A\oplus\Delta}(\mathbf{x})
=
\phi_A(\mathbf{x})
+
\pi\sum_{i<j}\Delta_{ij}x_i x_j
\quad (\mathrm{mod}\;2\pi)
\end{equation}
therefore maps the encoded adjacency matrix from $A$ to
$A\oplus\Delta$, while leaving the local phases $\vec{\theta}$
unchanged. Here, $\oplus$ denotes element-wise addition modulo two.

\begin{figure}
    \centering
    \includegraphics[width=0.9\columnwidth]{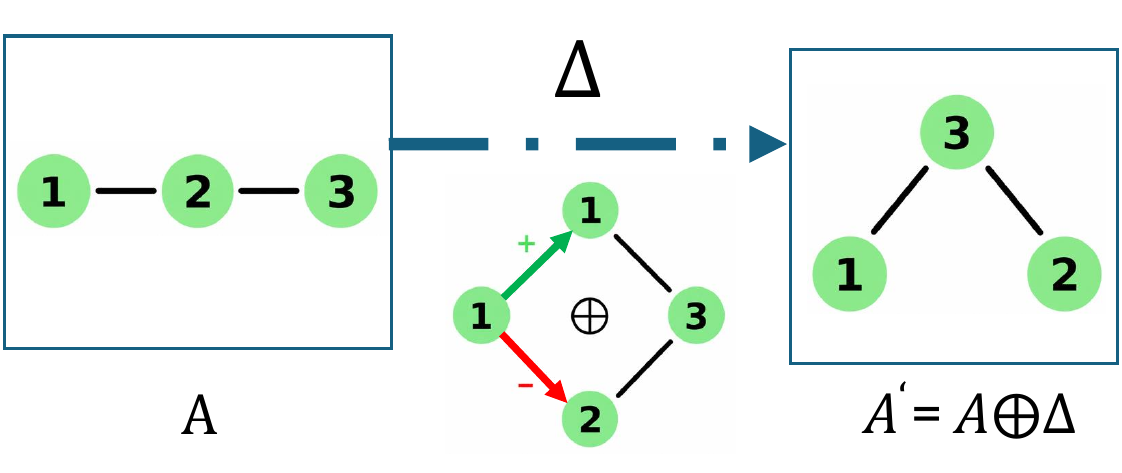}
    \caption{
    Graph-to-graph transformation through phase-function updates.
    A prescribed edge-toggle pattern $\Delta$ adds the quadratic
    phase $\pi\sum_{i<j}\Delta_{ij}x_i x_j$, mapping the encoded
    graph $A$ to $A'=A\oplus\Delta$ while preserving the local
    phases. Implementing this update requires only $\Delta$,
    whereas determining the pattern for a specified target $A'$
    requires the adjacency difference $A\oplus A'$.
    Sequential client contributions compose by addition modulo
    two. See Appendix~\ref{app:g2g} for details.
    }
    \label{fig:graph_transformation}
\end{figure}

For a specified initial graph $A$ and target graph $A'$, the
required toggle pattern is:
\begin{equation}
\Delta = A\oplus A'.
\end{equation}
Determining this pattern requires sufficient information to
identify the adjacency difference. Once $\Delta$ is specified,
however, implementing its phase update requires no knowledge
of $A$ or $\vec{\theta}$ (Fig.~\ref{fig:graph_transformation}).

\begin{figure}[t]
    \centering
    \hspace*{-0.5cm}%
    \includegraphics[width=0.9\columnwidth]{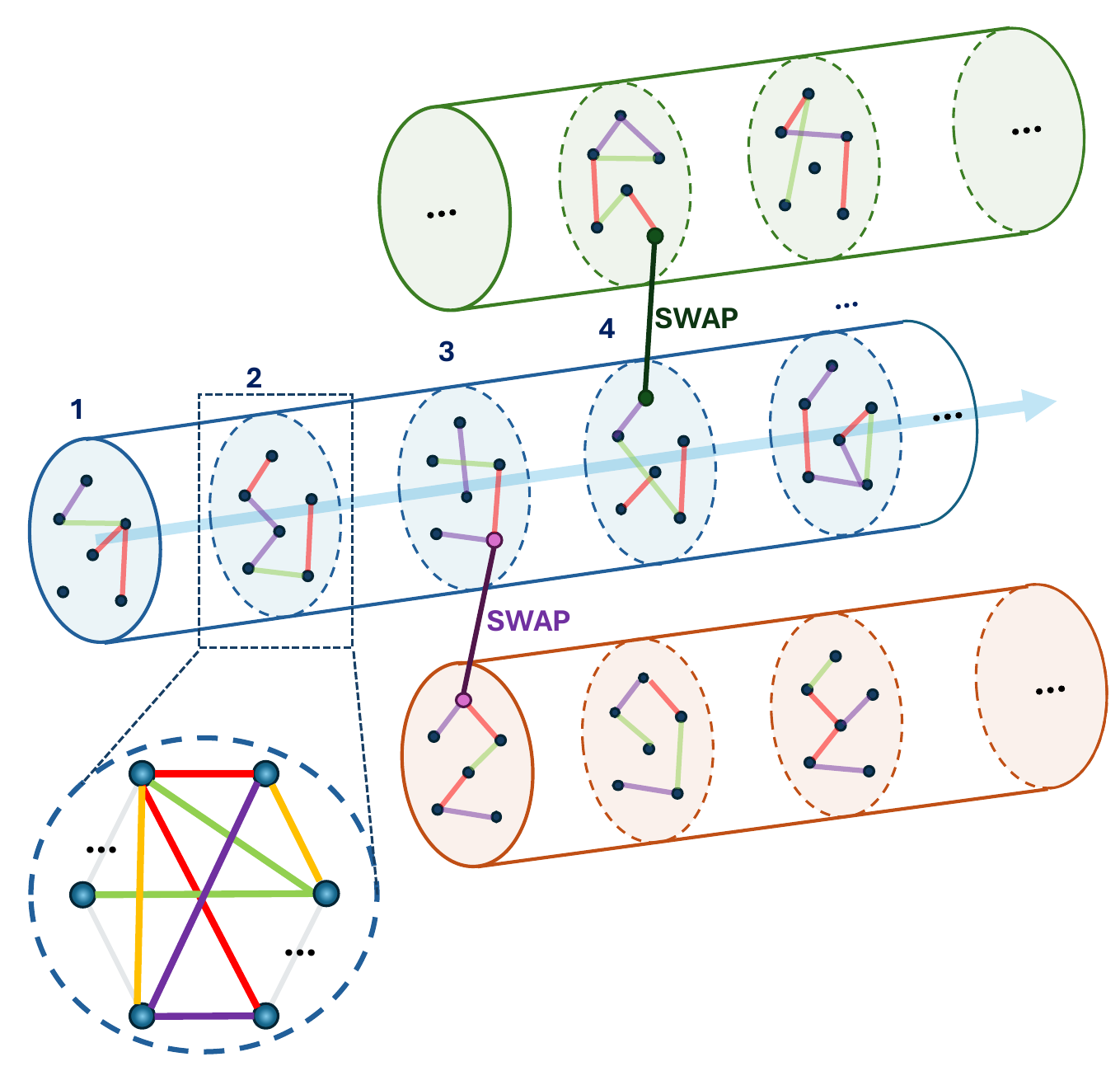}
    \caption{Conceptual tube-graph representation of RGSP-assisted UBQC.
    Each cross-sectional slice represents one complete graph state prepared from a single optical pulse, while the longitudinal direction corresponds to the logical measurement depth with server-generated entanglement between slices within each tube. The central tube acts as an entanglement backbone, so that SWAP operations (or CZ gates) are required only to generate entanglement across tubes.
    }
    \label{fig:tube}
\end{figure}

\lettersection{Application: Reduced-SWAP Universal Blind Quantum Computation}
\label{applications}
Direct preparation of arbitrary graph states fundamentally changes how universal blind quantum computation (UBQC) is implemented~\cite{broadbentUniversalBlindQuantum2009a}. We propose a three-dimensional \emph{tube-graph model} to illustrate the architectural advantage of RGSP-assisted UBQC (Fig.~\ref{fig:tube}). In this model, each cross-sectional slice corresponds to one complete graph state prepared from a single optical pulse, while the longitudinal direction represents the logical measurement depth of the computation. Long-range interactions within each qudit slice are generated via phase encoding, completely eliminating the need for SWAP operations within the slice. While communication between separate temporal qudits still requires SWAP gates across the longitudinal communication backbone, the number of SWAP operations is substantially reduced. 

To quantify this resource advantage, we compare the qubit resources required to implement the $N$-qubit QFT~\cite{Barak2007}, which is a building block for blind quantum machine learning~\cite{Fleur2025BQML}. For a brickwork topology:
\begin{equation}
N_{\mathrm{RSP}} = (\text{brickwork depth}) \times 4 \times (\text{circuit width}),
\label{eq:brickwork_resource}
\end{equation}

with much of the brickwork depth coming from SWAP gates across non-adjacent qubits. For a six-qubit QFT ($\text{QFT}_6$), we use the MasonQ transpiler to find a minimal brickwork state of 420 remotely prepared qubits due to the heavy SWAP overhead. Using $6$-qubit RGSP to remove SWAP gates, the same QFT is executed using only 48 remotely prepared qubits. This small example illustrates a dramatic saving of nearly 90\% of the relevant quantum resource (Table~\ref{tab:resource_comparison_appendix} and Fig.~\ref{fig:qft_rsp_resource_comparison}).

\begin{figure}[htbp]
    \centering
    \includegraphics[width=0.9\columnwidth]{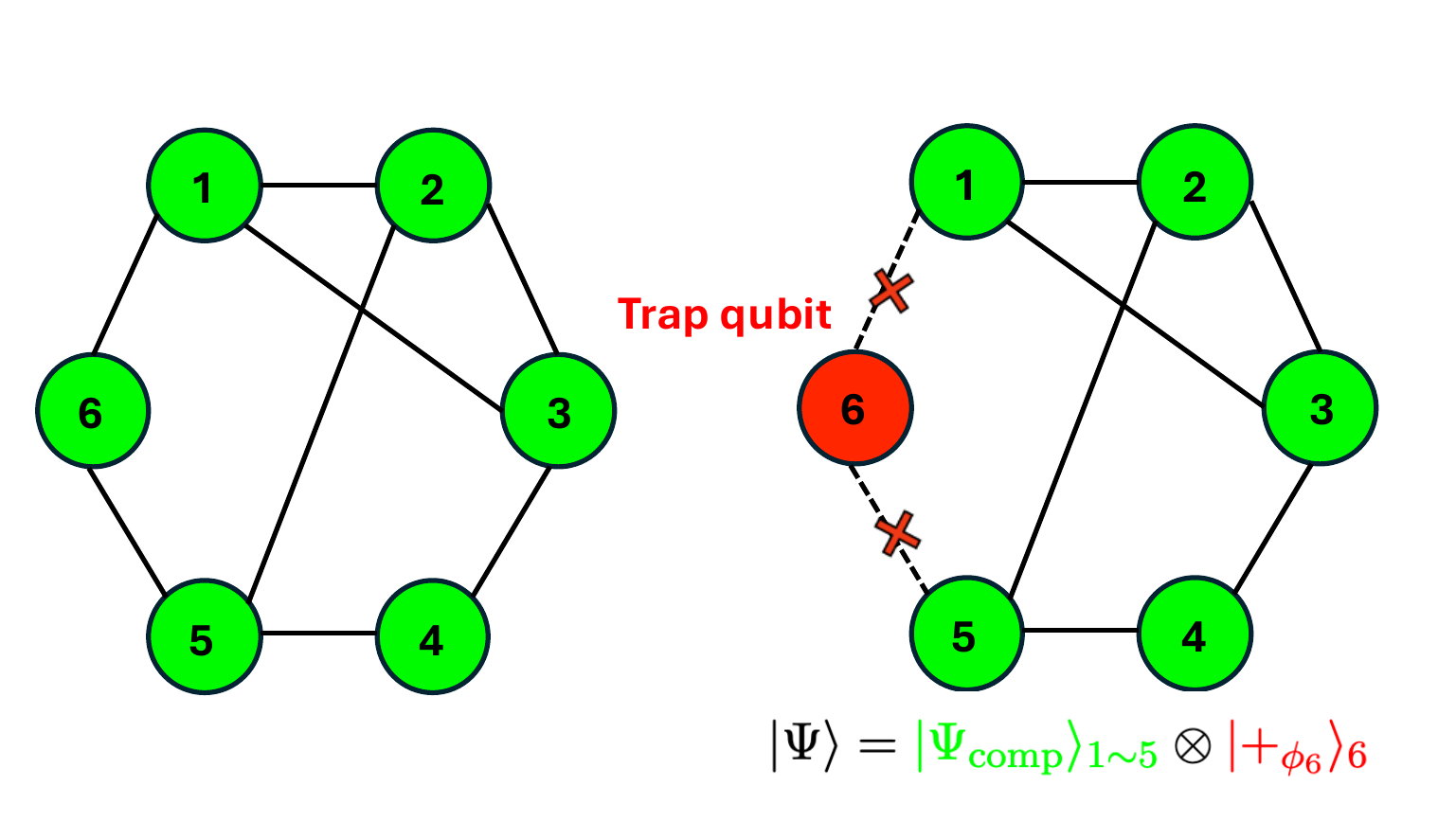}
    \caption{
    Direct preparation of a graph state containing an isolated trap qubit.
    Qubit~6 remains in the state $|+_{\theta_6}\rangle$ while the remaining qubits form the computational graph.}
    \label{fig:direct_trap_qubit}
\end{figure}


\paragraph*{Trap-based verification~\cite{Fitzsimons2017, kapourniotis_unifying_2024}.}
\label{para:trap}
Traditional UBQC uses dummy qubits to isolate selected trap locations from the computational resource to verify that the server faithfully implements the protocol. Since RGSP allows the client to directly prepare arbitrary graph states, trap qubits can be incorporated directly into the target graph. To isolate a trap qubit $i \in \{1,\dots,n\}$ at a secret angle $\theta_i$, the client simply sets the adjacency matrix elements to $A_{ij} = A_{ji} = 0$ for all $j \neq i$. This decouples the trap vertex from the rest of the graph, causing the multi-qubit phase function to factorize as:
\begin{equation}
\phi_x = \phi_{x_{\text{rest}}}^{(\text{comp})} + \theta_i x_i \quad (\mathrm{mod}\;2\pi),
\label{eq:trap_factorization}
\end{equation}
where $x_{\text{rest}}$ denotes the remaining bits. Consequently, the prepared state on the server factorizes into a product state:
\begin{equation}
|\Psi_G\rangle = |\Psi_{\text{comp}}\rangle \otimes |+_{\theta_i}\rangle_i .
\label{eq:trap_state_factorization}
\end{equation}
The server remains completely blind to the trap's presence because all temporal modes propagate through the same physical setup~\cite{broadbentUniversalBlindQuantum2009a,Fitzsimons2017}(Fig.~\ref{fig:direct_trap_qubit}), with performance agnostic to the graph topology. The client can then instruct the server to measure the trap in the $B_{\theta_i} = \{|+_{\theta_i}\rangle, |-_{\theta_i}\rangle\}$ basis. For an honest server, the decoded outcome $s_i = b_i \oplus r_i$ is deterministically $0$, providing a simple verification witness.

\paragraph*{Applications beyond blind quantum computation.}
Beyond UBQC, the ability to directly prepare arbitrary graph states may also benefit distributed graph-state protocols, including quantum position verification (QPV) and multi-party delegated quantum computation.

For QPV, encoding many logical qubits into a single transmitted photon is particularly attractive for protocols requiring multiple qubits to arrive within a limited relativistic time window, such as in parallel QPV~\cite{escolàfarràs2025quantumpositionverificationshot} or moving-prover scenarios like the Quantum Boat protocol~\cite{bartusek2026trackqubitsspacetime}. Whereas independent transmission succeeds with probability $\eta_t^n$, RGSP distributes all logical qubits within one photon, giving an overall transmission probability proportional to $\eta_t$. This substantially improves the transmission success probability of many-qubit graph-state resources while remaining compatible with relativistic timing constraints (Appendix~\ref{app:qpv}).

RGSP also provides a graph-state preparation primitive for multi-party delegated quantum computation. A multi-client QLine architecture has already been demonstrated experimentally, with proposed applications such as federated quantum machine learning~\cite{polacchiMulticlientDistributedBlind2023,
liQuantumFederatedLearning2021}. RGSP provides a complementary approach by enabling sequential phase accumulation on a single propagating qudit within such a QLine architecture, with amenability to near-term all-optical implementation as the technology for many-qubit quantum servers matures~\cite{polacchiMulticlientDistributedBlind2023}
(Fig.~\ref{fig:qline_grsp_architecture}).

\lettersection{Conclusions}
RGSP offers a scalable and noise-robust route toward near-term delegated quantum processors to perform quantum-secured delegated tasks, freeing UBQC protocols from the brickwork topology and easing implementation of traps for verification. While simultaneous preparation of an entire graph (or hypergraph \cite{Rossi2013}) for a large computation is prohibited by the exponential scaling in temporal modes (emphatically, there is no free lunch), sub-graphs can be glued together in the tube-graph framework we propose. For small computations, the removal of server-side multi-qubit gates eases the implementation of near-term UBQC experiments. Furthermore, the native client operations (including single qubit rotations and graph-to-graph transformations) provide a framework both for multi-client protocols for distributed computation in the style of QLine as well as QPV, and possibly more protocols waiting to be discovered. We hope that RGSP, by enabling a new method of delegated computation, accelerates the development of quantum-network protocols and applications. 

\lettersection{Acknowledgements}
We are grateful to Florian Wiesner for helpful conversations about this project.
TzBP gratefully acknowledges support from the Quantum Software Consortium Ada Lovelace Fellowship. This work was funded by the European Union's Horizon Europe research and innovation programme under grant agreement No.~101102140 – QIA Phase 1.
  BT acknowledge the support of Danmarks Grundforskningsfond (DNRF Grant No.~139, Hy-Q Center for Hybrid Quantum Networks).

\newpage
\bibliography{references_PRL_standardized}
\appendix

\newpage
\section{Derivation of the graph-state phase-function representation} 
\label{app: derivation_of_RGSP}

A graph state $|G\rangle$ associated with an undirected, simple graph $G=(V,E)$ on $n$ vertices is conventionally defined by applying controlled-phase ($CZ$) gates to a uniform product state superposition of all qubits initialized in the $|+\rangle$ state~\cite{rimock_generalized_2026, west2001introduction}:
\begin{equation}
|G\rangle
=
\prod_{(i,j)\in E}
CZ_{ij}
|+\rangle^{\otimes n},
\label{eq:graphstate_def_appendix}
\end{equation}
where $|+\rangle=\frac{|0\rangle+|1\rangle}{\sqrt2}$.

The initial unentangled state $|+\rangle^{\otimes n}$ can be expanded in the computational basis as:
\begin{equation}
|+\rangle^{\otimes n} = \frac{1}{2^{n/2}} \sum_{\vec{x} \in \{0,1\}^n} |\vec{x}\rangle,
\label{eq:uniform_superposition_appendix}
\end{equation}
where $|\vec{x}\rangle = |x_1 x_2 \dots x_n\rangle$ is the computational basis state corresponding to the $n$-bit string $\vec{x}$.

Since the $CZ_{ij}$ gate acts on the computational basis as $CZ_{ij}|x_i x_j\rangle = (-1)^{x_i x_j} |x_i x_j\rangle$, the application of all gates corresponding to the edge set $E$ yields:
\begin{align}
    |G\rangle &= \frac{1}{2^{n/2}} \sum_{\vec{x} \in \{0,1\}^n} (-1)^{\sum_{(i,j)\in E} x_i x_j} |\vec{x}\rangle\\
    &= \frac{1}{2^{n/2}} \sum_{\vec{x} \in \{0,1\}^n} e^{i \pi \sum_{i<j} A_{ij} x_i x_j} |\vec{x}\rangle,
\label{eq:entangled_graph_state_appendix}
\end{align}
where $A$ is the symmetric $n \times n$ binary adjacency matrix of the graph $G$, with $A_{ij} = 1$ if $(i,j)\in E$ and $A_{ij}=0$ otherwise.

To prepare graph states for MBQC, the phases $\vec{\theta} = (\theta_1, \dots, \theta_{n})$ are assigned to each qubit, introducing the linear phase contribution $\exp\left(i\sum_{i=1}^{n}\theta_i x_i\right)$.

Combining the local phases with the graph-dependent quadratic phase yields~\cite{propp2026remotely}:

\begin{equation}
|G(\vec{\theta})\rangle 
=
\frac{1}{2^{n/2}}
\sum_{\vec{x} \in \{0,1\}^n}
e^{i \phi_x}
|\vec{x}\rangle,
\label{eq:phase-function}
\end{equation}
where

\begin{equation}
\phi_x := \sum_{i=1}^{n} \theta_i x_i + \pi \sum_{i<j} A_{ij} x_i x_j.
\label{eq:phase_function_appendix}
\end{equation}

Here, the linear term specifies the local single-qubit rotations and the quadratic term determines the graph topology through the adjacency matrix. Eq.~(\ref{eq:phase_function_appendix}) is identical to the phase function introduced in Eq.~\eqref{eq:phase-rule}.

\section{RGSP protocol} 
\label{app:RGSP_protocol}
We describe the physical implementation of RGSP, which realizes the phase-function representation
derived in Appendix~\ref{app: derivation_of_RGSP} using a single
high-dimensional photonic qudit. The overall protocol is illustrated in
Fig.~\ref{fig:overview-grsp-bqc}.

\paragraph*{Encoded photonic state}
The client specifies the target graph $G=(V,E)$ and the local preparation angles
$\vec{\theta}=(\theta_1,\dots,\theta_n)$.
In a BQC application, these preparation angles are combined with the computation angles, classical randomization, and adaptive corrections when determining the subsequent
measurement instructions. The client computes the complete phase function $\phi_x$ according to Eq.~\eqref{eq:phase-rule} and prepares a single photon in a coherent superposition over $d=2^n$ temporal (or spatial) modes~\cite{Salakhutdinov2012,Hiesmayr2016,yu_two-photon_2015,kiktenkoColloquiumQuditsDecomposing2025}:
\begin{equation}
    |\psi_c\rangle = \sum_{\vec{x} \in \{0,1\}^n} c_x e^{i \phi_x} a_x^\dagger |\text{vac}\rangle,
\label{eq:prepared_photon_appendix}
\end{equation}

Here, $a_x^\dagger$ creates a single photon in mode $x$, and $\sum_x |c_x|^2=1$. For ideal lossless preparation, $c_x=2^{-n/2}$.
If mode-dependent attenuation is present, the input amplitudes can instead be chosen to compensate for the corresponding transmission amplitudes, so that the successfully transmitted state has equal mode amplitudes.

\paragraph*{Light-matter interaction}
The server initializes its $n$ memory qubits (such as trapped ions or solid-state spins) in
$|+\rangle^{\otimes n}$. Conditioned on the photonic mode
label $x$, the light--matter interaction applies
$\bigotimes_{i=1}^{n}Z_i^{x_i}$ to the register.
Applying $H^{\otimes n}$ to the memories then maps:
\begin{equation}
H^{\otimes n}
\left(\bigotimes_{i=1}^{n}Z_i^{x_i}\right)
|+\rangle^{\otimes n}
=
|\vec{x}\rangle_s.
\end{equation}

Thus, the joint state~\cite{propp2026remotely}:
    \begin{equation}
    |\Phi\rangle = \sum_{\vec{x} \in \{0,1\}^n} c_x e^{i \phi_x} a_x^\dagger |\text{vac}\rangle \otimes |\vec{x}\rangle_s,
    \label{eq:joint_entangled_state_appendix}
    \end{equation}
    where $|\vec{x}\rangle_s$ denotes the computational basis state of the server's memory register.

\paragraph*{Which-mode erasure \& heralded projection}
To project the memory register into the target entangled graph state, the photonic modes are mixed in a multi-port interferometric network (e.g., a time-lens or a spectral QFT network) implementing $H^{\otimes n}$ to erase the path and time-bin information. Conditioned on a single-photon detection click at the output, the which-mode information is completely erased, projecting the remote memories into the state:
\begin{equation}
    |\Psi\rangle = \frac{1}{2^{n/2}} \hspace{-3mm} \sum_{\vec{x} \in \{0,1\}^n} \hspace{-3mm} \exp\hspace{-1mm}\left[ i \hspace{-0.5mm}\left(\hspace{-0.5mm} \sum_{i=1}^n \theta_i x_i + \pi \sum_{i<j} A_{ij} x_i x_j \hspace{-1mm}\right) \hspace{-1mm}\right] |\vec{x}\rangle_s.
\label{eq:recovered_graph_state_appendix}
\end{equation}
Since $e^{i \pi x_i x_j} = (-1)^{x_i x_j}$, this state is mathematically equivalent to $|G(\vec{\theta})\rangle$, which is the desired entangled graph state on the server.

\paragraph*{Integration with MBQC}

After successful heralding, the prepared graph state is consumed as
the computational resource for MBQC. The server performs the prescribed single-qubit measurements and returns the measurement outcomes $b_k\in\{0,1\}$. The client uses these outcomes together with the classical
description of the computation to determine the adaptive measurement
angles for subsequent qubits. 

\subsection*{Example: Two-Qubit State Preparation with One Photon}
\label{app:derivation_grsp}
To illustrate the phase-encoding mechanism of RGSP, we analyze the explicit preparation of a two-qubit graph state ($n=2$) using a single photon distributed across $d=2^2=4$ optical modes.

The client prepares a single photon in a coherent superposition over modes labelled by bitstrings \(\vec{x}=(x_1,x_2)\in\{0,1\}^2\),
\begin{equation}
|\psi_c\rangle = \sum_{\vec{x}\in\{0,1\}^2} c_x e^{i\phi_x} a_x^\dagger |\mathrm{vac}\rangle,
\label{eq:app_client_state_general}
\end{equation}
where \(a_x^\dagger\) creates a photon in mode \(x\). For equal mode amplitudes ($c_x = 1/2$), the state expands as:

\begin{equation}
\begin{aligned}
|\psi_c\rangle = \frac{1}{2}\bigl(
&e^{i\phi_{00}}|1_{00}\rangle
+e^{i\phi_{01}}|1_{01}\rangle \\
&+e^{i\phi_{10}}|1_{10}\rangle
+e^{i\phi_{11}}|1_{11}\rangle
\bigr).
\end{aligned}
\label{eq:app_client_equal_superposition}
\end{equation}

The server initializes two memory qubits in \(|+\rangle^{\otimes 2}\). Conditioned on the photonic mode label \(\vec{x}=(x_1,x_2)\), the photon induces phase flips on the corresponding memory qubits:
\begin{equation}
|1\rangle_x |+\rangle^{\otimes 2} \longrightarrow |1\rangle_x Z_1^{x_1}Z_2^{x_2}|+\rangle^{\otimes 2}.
\label{eq:app_conditional_z_action}
\end{equation}

Since \(Z|+\rangle=|-\rangle\), the joint photon--memory state becomes:
\begin{equation}
\begin{aligned}
|\Psi\rangle = \frac{1}{2}\bigl(
&e^{i\phi_{00}}|1_{00}\rangle|++\rangle
+e^{i\phi_{01}}|1_{01}\rangle|+-\rangle \\
&+e^{i\phi_{10}}|1_{10}\rangle|-+\rangle
+e^{i\phi_{11}}|1_{11}\rangle|--\rangle
\bigr).
\end{aligned}
\label{eq:app_after_interaction_pm}
\end{equation}

Applying Hadamard gates to the memory qubits maps \(H|+\rangle=|0\rangle\) and \(H|-\rangle=|1\rangle\). Hence, the joint photon-matter state becomes:
\begin{equation}
\begin{aligned}
|\Psi\rangle = \frac{1}{2}\bigl(
&e^{i\phi_{00}}|1_{00}\rangle|00\rangle
+e^{i\phi_{01}}|1_{01}\rangle|01\rangle \\
&+e^{i\phi_{10}}|1_{10}\rangle|10\rangle
+e^{i\phi_{11}}|1_{11}\rangle|11\rangle
\bigr).
\end{aligned}
\label{eq:app_joint_correlated_state}
\end{equation}

Applying the erasure operator $|\chi\rangle = \frac{1}{2}\sum_{\vec{x}\in\{0,1\}^2} |1\rangle_x$ projects the optical modes onto a uniform superposition, leaving the remote memories in the unnormalized state:
\begin{equation}
\langle \chi | \Psi \rangle = \frac{1}{4} \left( e^{i \phi_{00}}|00\rangle + e^{i \phi_{01}}|01\rangle + e^{i \phi_{10}}|10\rangle + e^{i \phi_{11}}|11\rangle \right).
\label{eq:two_qubit_projected_appendix}
\end{equation}

Normalizing the register state gives:
\begin{equation}
|\Psi_s\rangle = \frac{1}{2} \left( e^{i \phi_{00}}|00\rangle + e^{i \phi_{01}}|01\rangle + e^{i \phi_{10}}|10\rangle + e^{i \phi_{11}}|11\rangle \right).
\label{eq:two_qubit_normalized_appendix}
\end{equation}

To prepare the entangled target state $|G\rangle = CZ_{12} |+_{\theta_1}\rangle |+_{\theta_2}\rangle$, we program the phase function using the quadratic rule of Eq.~\eqref{eq:phase-rule}:
\begin{equation}
\phi_x = \theta_1 x_1 + \theta_2 x_2 + \pi x_1 x_2.
\label{eq:two_qubit_phase_choice_appendix}
\end{equation}
This prescribes the optical phases:
\begin{equation}
\phi_{00} = 0, \hspace{2mm} \phi_{01} = \theta_2, \hspace{2mm} \phi_{10} = \theta_1, \hspace{2mm} \phi_{11} = \theta_1 + \theta_2 + \pi.
\label{eq:two_qubit_phases_appendix}
\end{equation}

Substituting these phases back into Eq.~\eqref{eq:two_qubit_normalized_appendix} yields:
\begin{equation}
|\Psi_s\rangle = \frac{1}{2} \left( |00\rangle + e^{i \theta_2}|01\rangle + e^{i \theta_1}|10\rangle - e^{i (\theta_1 + \theta_2)}|11\rangle \hspace{-0.5mm}\right),
\label{eq:two_qubit_recovered_appendix}
\end{equation}


which is exactly the target state
$CZ_{12}|+_{\theta_1}\rangle|+_{\theta_2}\rangle$.
The preparation therefore requires no direct entangling
gates between the server's memory qubits; entanglement is generated through the photon--memory interactions and the subsequent photonic measurement.

Reflection-based protocols have previously been
investigated for entanglement generation
\cite{nemoto-2014-photon_archit_scalab_quant_infor_proces_diamon,Omlor2025,Zheng2022}.
Building on this line of work, the present construction uses the quadratic phase $\pi\sum_{i<j}A_{ij}x_i x_j$
to specify the graph edges, while the linear phase terms specify the local rotations. The example above illustrates how this programmed phase structure is transferred to the remote memory register.

\section{Proof of the topology-independent fidelity}
\label{app:fidelity}
We evaluate the impact of channel phase errors on the state prepared at the server register. The ideal target state prepared by the client is:
\begin{equation}
|\psi_{\mathrm{ideal}}\rangle = \sum_{\vec{x}\in\{0,1\}^n} c_x e^{i\phi_x}|\vec{x}\rangle,
\label{eq:ideal_state}
\end{equation}

Optical phase noise in the fiber channel introduces a stochastic phase error $\epsilon_x$ on each temporal mode $x$ during transmission, so that the actual state generated on the server is:
\begin{equation}
|\psi_{\mathrm{noisy}}\rangle = \sum_{\vec{x}\in\{0,1\}^n} c_x e^{i(\phi_x+\epsilon_x)}|\vec{x}\rangle.
\label{eq:noisy_state}
\end{equation}

The fidelity of the prepared state relative to the ideal state is defined by the overlap $F = |\langle \psi_{\text{ideal}} | \psi_{\text{noisy}} \rangle|^2$:
\begin{equation}
\begin{aligned}
\langle \psi_{\text{ideal}} | \psi_{\text{noisy}} \rangle
&= \sum_{\vec{x} \in \{0,1\}^n}
|c_x|^2 e^{-i\phi_x} e^{i(\phi_x+\epsilon_x)} \\
&= \sum_{\vec{x} \in \{0,1\}^n}
w_x e^{i\epsilon_x}.
\end{aligned}
\label{eq:overlap_derivation_appendix}
\end{equation}
where $w_x = c_x^2 = |c_x|^2$ represents the normalized mode weights.

The fidelity $F$ is therefore:
\begin{align}
F = \left| \sum_{\vec{x} \in \{0,1\}^n} w_x e^{i \epsilon_x} \right|^2.
\label{eq:fidelity_result_appendix}
\end{align}

A fundamental physical insight is that the graph-dependent phase function $\phi_x$ of Eq.~\eqref{eq:phase-rule} cancels out completely in the overlap derivation in Eq.~\eqref{eq:overlap_derivation_appendix}. Consequently, the state fidelity depends strictly on the channel phase noise $\epsilon_x$ and is fundamentally independent of the graph topology. 

Furthermore, because Eq.~\eqref{eq:fidelity_result_appendix} depends only on the relative phase differences between temporal modes, the protocol is immune to common-mode phase fluctuations. If a global phase drift $\epsilon_0$ is added to all modes ($\epsilon_x \to \epsilon_x + \epsilon_0$), the global phase factor $e^{i \epsilon_0}$ factors out of the sum and is eliminated by the absolute value square, so that:
\begin{equation}
F = \left| e^{i \epsilon_0} \sum_{\vec{x}} w_x e^{i \epsilon_x} \right|^2 = \left| \sum_{\vec{x}} w_x e^{i \epsilon_x} \right|^2.
\label{eq:common_mode_cancellation_appendix}
\end{equation}
Thus differential phase sensitivity demands only relative phase stability within a single qudit pulse train, rather than sub-wavelength absolute phase stabilization between remote nodes.

\section{Details on the noise models}
\label{app:noise}
The fidelity analysis in the main text assumes that the dominant imperfections arise from optical phase fluctuations accumulated during the preparation and transmission of the encoded photonic pulse train. In this appendix we define the phase-noise models used throughout the numerical simulations and discuss the corresponding pulse-ordering strategy. Throughout this work we assume that amplitude loss is compensated independently through the mode amplitudes $c_x$, while all decoherence enters through the stochastic phase variables $\epsilon_x$. 

\subsection{Independent phase-noise model}
The first model assumes that every temporal mode experiences an independent Gaussian phase error, the phase is:
\begin{equation}
\phi_x \;\longrightarrow\; \phi_x + \epsilon_x(L),
\end{equation}
where $\epsilon_x(L)$ denotes the fibre-induced phase noise,
\begin{equation}
\epsilon_x(L) = \frac{2\pi}{\lambda} \int_0^L \delta n_x(z) \, dz,
\label{eq:phase_integral}
\end{equation}
with $\delta n_x(z)$ being the local refractive-index fluctuation along the fibre, which means the phase arises directly from an accumulated path integral of refractive-index variations.

In practice, the fibre is subject to a large number of independent perturbations e.g., thermal fluctuations, mechanical vibrations, and air currents. By the \textit{Central Limit Theorem}, the integrated effect in Eq.~\eqref{eq:phase_integral} can be approximated as a normal distribution:
\begin{equation}
\epsilon_x(L) \;\sim\; \mathcal{N}\bigl(0,\, \sigma_\phi^2(L)\bigr),
\end{equation}
where $\sigma_\phi^2(L)$ is the phase variance and $L$ is the fibre distance.

The expected fidelity is:
\begin{equation}
F_{\mathrm{iid}}
=
\left| \frac{1}{N} \sum_{x} e^{i \epsilon_x} \right|^2.
\end{equation}

After averaging, 

\begin{equation}
\mathbb{E}[F_{\text{iid}}] = \frac{1}{N} + \left(1 - \frac{1}{N}\right) e^{-\sigma_\phi^2(L)}.
\label{eq:iid_phase_noise_expected_fidelity}
\end{equation}

Take the Gaussian phase diffusion model proposed by Minář~\cite{Minar2007}, which is $\sigma_{\phi}^{2}(L) = 2 D_{\phi} L$, where $D_{\phi}$ is the phase diffusion coefficient. Consequently, the fidelity as a function of distance is:

\begin{equation}
\mathbb{E}[F_{\text{iid}}] = \frac{1}{N} + \left(1 - \frac{1}{N}\right) e^{-2 D_{\phi} L}.
\end{equation}

This model describes situations in which active phase stabilization suppresses long-time correlations, so that different temporal modes accumulate statistically independent phase errors.

\subsection{Cumulative phase-noise model}

To describe slow environmental drifts in long optical fibers, we additionally consider a cumulative phase-noise model, equivalently as a Wiener process (Brownian motion), where the phase error of the $k$-th temporal mode accumulates all preceding fluctuations. Ordering modes chronologically as \(x_{1}, x_{2}, \dots, x_{N}\) (\(N = 2^{n}\)), the phase error on the \(k\)-th mode is:

\begin{equation}
\epsilon_{x_k}
=
\sum_{\ell \leq k} \xi_{\ell},
\qquad
\xi_{\ell} \sim \mathcal{N}(0,\sigma_\phi^2(L)\bigr).
\label{eq:cumulative_phase_error}
\end{equation}
with independent increments \(\xi_{\ell}\). The fidelity then becomes:
\begin{equation}
F_{\mathrm{cum}}
=
\left|
\sum_{k=1}^{N}
w_{x_k} e^{i\epsilon_{x_k}} \right|^2,
\end{equation}

so that
\begin{equation}
F_{\mathrm{cum}}
=
\sum_{k,l}
w_{x_k}w_{x_l}
e^{i(\epsilon_{x_k}-\epsilon_{x_l})}.
\end{equation}

For a random walk, the phase difference satisfies:
\begin{equation}
\epsilon_{x_k}-\epsilon_{x_l}
\sim
\mathcal{N}
\left(
0,
|k-l| \sigma_\phi^2(L)
\right),
\end{equation}

Taking the expectation,
\begin{equation}
\mathbb{E}
\left[
e^{i(\epsilon_{x_k}-\epsilon_{x_l})}
\right]
=
e^{-\frac{1}{2}|k-l|\sigma_\phi^2(L)}.
\end{equation}

Thus, the expected fidelity is:
\begin{equation}
\mathbb{E}[F_{\mathrm{cum}}]
=
\sum_{k,l}
w_{x_k}w_{x_l}
e^{-\frac{1}{2}|k-l|\sigma_\phi^2(L)}.
\label{eq:cum_weighted_fidelity}
\end{equation}

\begin{figure}
    \centering
    \hspace*{-0.4cm}
    \includegraphics[width=1.1\columnwidth]{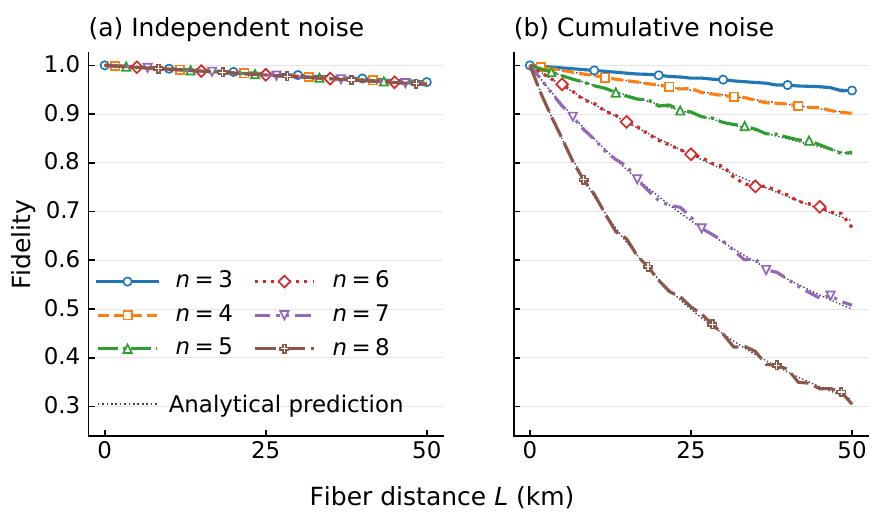}
    \caption{Output qubit graph-state fidelity for brickwork states with uniform temporal-mode amplitudes. \textit{Left:} Independent phase noise, where fidelity decays weakly with \(n\), as each mode experiences the same variance independently. \textit{Right:} Cumulative phase noise, where later modes accumulate drift from all previous steps. This induces strong correlations and rapid fidelity decay with both $\epsilon_x$ and \(n\). Since the number of temporal modes scales as \(2^n\), larger graph states are highly sensitive to cumulative fiber phase drift.}
    \label{fig:n_behavior}
\end{figure}

Unlike the independent model, neighboring temporal modes are therefore correlated, and the accumulated phase uncertainty increases monotonically along the pulse train, as illustrated in Fig.~\ref{fig:n_behavior}. 

\subsection{Mode-dependent amplitudes}
During transmission through the light-matter interface, a mode $x$ undergoes interactions with $i$th element of the matter register if and only if the corresponding binary digit $x_i = 1$. The transmission efficiency for mode $x$ depends exponentially on its Hamming weight $H(x)$, which counts the number of active light-matter gate interactions~\cite{Rethinasamy2024}:
\begin{equation}
    \eta_x
    =
    \eta_t\eta_d
    \eta_1^{H(x)}
    \eta_0^{n-H(x)} .
    \label{eq:mode_dependent_efficiency}
\end{equation}

where $\eta_t$ and $\eta_d$ are the baseline transmission and detector efficiencies, and $\eta_1$ ($\eta_0$) represents the mode efficiency for interacting (non-interacting) modes.

\begin{figure}
    \centering
    \hspace*{-2.0cm}
    \includegraphics[width=0.7\columnwidth]{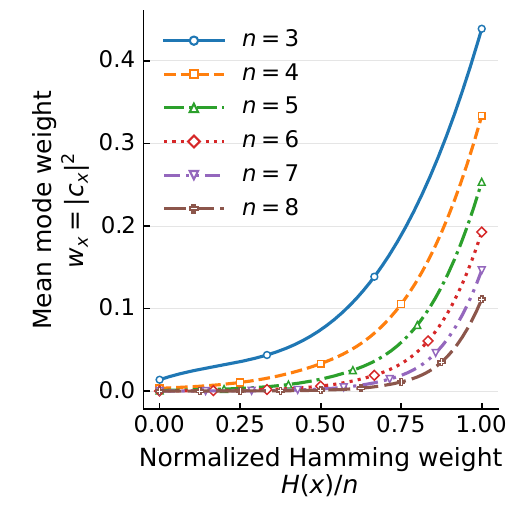}
    \caption{ Average loss-compensated mode weight \(w_x=|c_x|^2\) as a function of the normalized Hamming weight \(H(x)/n\) for \(n=3\) to \(n=8\).
    The weight grows rapidly with Hamming weight, reflecting the stronger attenuation of high-weight modes. 
    This effect becomes more pronounced for larger systems because the range of accessible Hamming weights expands with \(n\), leading to a broader distribution of transmission efficiencies.
    }
    \label{fig:loss_compensated_weights}
\end{figure}

To prepare a state with uniform target amplitudes $c_{\text{target}} = 1/2^{n/2}$, the client must pre-compensate the mode-dependent attenuation at preparation by shaping the initial amplitudes as~\cite{propp2026remotely}:
\begin{equation}
c_x \propto \frac{1}{\sqrt{\eta_x}}.
\label{eq:pre_compensation_appendix}
\end{equation}
This pre-compensation concentrates the normalized mode weights $w_x = |c_x|^2$ onto high-Hamming-weight modes (as illustrated in Fig.~\ref{fig:loss_compensated_weights}).

\begin{figure*}
    \centering
    \includegraphics[width=0.9\textwidth]{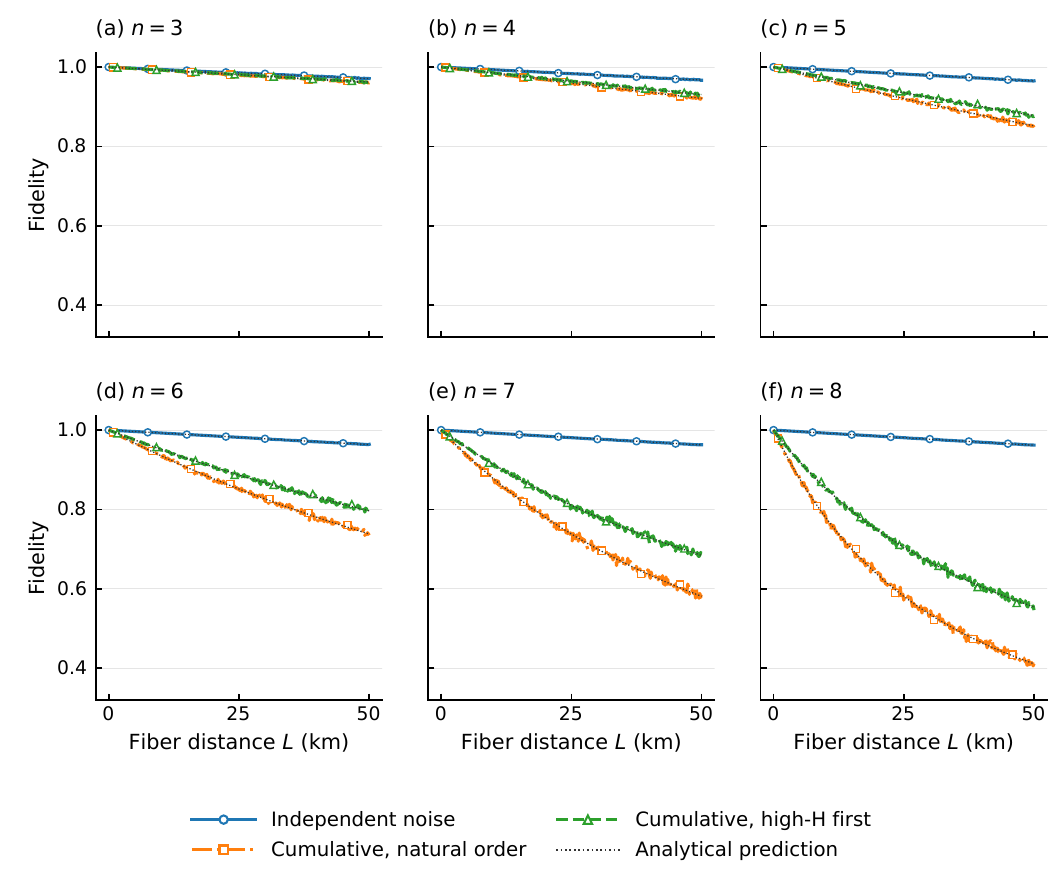}
    \caption{Fidelity as a function of distance for different graph sizes and noise models. Ordering comparison for loss-compensated brickwork states with \(n=3\) to \(n=8\).
    The blue curves show the independent-noise baseline, the orange curves show cumulative phase noise with natural ordering, and the green curves show cumulative phase noise with high-Hamming-weight-first ordering.
    The dashed curves show the corresponding analytical predictions.
    The benefit of reordering becomes more pronounced as \(n\) increases, because loss compensation concentrates more weight on high-Hamming-weight modes.
    Placing these modes earlier in the temporal sequence reduces the accumulated phase drift affecting the most important components of the state.}
    \label{fig:ordering_comparison_loss_comp_distance}
\end{figure*}

These weights enter directly into the fidelity expression Eqs.~\eqref{eq:iid_phase_noise_expected_fidelity} and \eqref{eq:cum_weighted_fidelity}. Because loss compensation assigns exponentially larger weights $w_x$ to high-Hamming-weight modes, these modes contribute most dominantly to the overall state fidelity in Eq.~\eqref{eq:fidelity_result_appendix}. To minimize their phase drift, the client can employ a mode reordering strategy wherein high-Hamming-weight modes (i.e. those that contribute most to the final state fidelity as they interact with the most qubits) are placed first in the pulse sequence ("High-H First"). This reordering minimizes the cumulative phase drift for the most heavily weighted modes, significantly improving the overall state fidelity (as shown in Fig.~\ref{fig:ordering_comparison_loss_comp_distance}).

\section{Details on graph-to-graph transformations} 
\label{app:g2g}
Let $\mathcal{G}_n$ be the set of simple undirected graphs on
$n$ labelled vertices, identified with symmetric $n\times n$
binary adjacency matrices with vanishing diagonal. An
edge-toggle pattern is specified by a matrix $\Delta$ with
the same symmetry and diagonal constraints. It defines the map:
\begin{equation}
T_\Delta:\mathcal{G}_n\to\mathcal{G}_n,
\qquad
T_\Delta(A)=A\oplus\Delta,
\end{equation}
where $\oplus$ denotes element-wise addition modulo two.
For each pair of vertices,
\begin{equation}
(A\oplus\Delta)_{ij}
=
\begin{cases}
1-A_{ij}, & \Delta_{ij}=1,\\
A_{ij},   & \Delta_{ij}=0.
\end{cases}
\label{eq:modulo_addition_cases_appendix}
\end{equation}
Thus, a toggle removes a present edge or adds an absent edge,
while entries with $\Delta_{ij}=0$ leave the graph unchanged.

\subsection*{Implementation through phase updates}

For fixed local phases $\vec{\theta}$, the phase function
encoding a graph $A$ is:
\begin{equation}
\phi_A(\vec{x})
=
\vec{\theta}^{\,T}\vec{x}
+
\pi\sum_{i<j}A_{ij}x_i x_j,
\qquad
\vec{x}\in\{0,1\}^n.
\end{equation}
Writing $A'=A\oplus\Delta$, the target phase function is:
\begin{equation}
\phi_{A'}(\vec{x})
=
\vec{\theta}^{\,T}\vec{x}
+
\pi\sum_{i<j}A'_{ij}x_i x_j.
\label{eq:target_phase_appendix}
\end{equation}
For binary entries,
\begin{equation}
A_{ij}\oplus\Delta_{ij}
=
A_{ij}+\Delta_{ij}-2A_{ij}\Delta_{ij}.
\end{equation}
The last term contributes only integer multiples of $2\pi$
to the phase, so
\begin{equation}
\phi_{A'}(\vec{x})
=
\phi_A(\vec{x})
+
\pi\sum_{i<j}\Delta_{ij}x_i x_j
\quad(\mathrm{mod}\;2\pi).
\label{eq:phase_transformation_appendix}
\end{equation}
This update is implemented on the photonic qudit by the
diagonal operation:
\begin{equation}
U_\Delta
=
\sum_{\vec{x}\in\{0,1\}^n}
\exp\!\left(
i\pi\sum_{i<j}\Delta_{ij}x_i x_j
\right)
|\vec{x}\rangle\langle\vec{x}|,
\label{eq:toggle_unitary_appendix}
\end{equation}
The operation changes the encoded adjacency matrix
while preserving the local phases $\vec{\theta}$.

For a specified initial graph $A$ and target graph $A'$, the
required pattern is $\Delta=A\oplus A'$. Determining this
pattern requires sufficient information to identify their
adjacency difference. Once $\Delta$ is specified, however,
implementing $U_\Delta$ requires no knowledge of $A$ or
$\vec{\theta}$. In particular, applying a prescribed toggle
pattern to an unknown graph does not require reconstructing
that graph.

\subsection*{Composition of multiple clients' contributions}

In the multi-client setting, client $k$ contributes its own
edge-toggle pattern $\Delta^{(k)}$ and applies
$U_{\Delta^{(k)}}$ to the same photonic qudit. The toggle
operations satisfy:
\begin{equation}
U_{\Delta^{(2)}}U_{\Delta^{(1)}}
=
U_{\Delta^{(1)}\oplus\Delta^{(2)}}.
\label{eq:toggle_composition_appendix}
\end{equation}
Consequently, after $M$ clients have acted,
\begin{equation}
U_{\mathrm{tot}}
=
U_{\Delta^{(M)}}\cdots U_{\Delta^{(1)}}
=
U_{\Delta_{\mathrm{tot}}},
\qquad
\Delta_{\mathrm{tot}}
=
\bigoplus_{k=1}^{M}\Delta^{(k)},
\end{equation}
and the resulting graph is:
\begin{equation}
A_{\mathrm{out}}
=
A_{\mathrm{in}}
\oplus
\bigoplus_{k=1}^{M}\Delta^{(k)}.
\label{eq:multiclient_graph_output_appendix}
\end{equation}
Thus, the final graph is determined \emph{jointly} by the initial
graph and the clients' contributions and it need not be specified
as a common target before the clients act.

To implement its update, each client needs only its own
pattern and the shared convention relating bit strings to
photonic modes. Knowledge of the intermediate graph, the
other clients' patterns, or the final graph is unnecessary.
No classical aggregation of these patterns is required for
the graph update itself since their contributions accumulate
\emph{directly} through the sequential phase operations. Classical
communication for subsequent measurement instructions and
outcome processing is a separate part of the computation
protocol.

For example, if two clients both toggle edge $(i,j)$, their
contributions cancel because $1\oplus1=0$, irrespective of
whether the edge was initially present. More generally,
the toggle operations commute, the zero pattern gives the
identity, and every operation is its own inverse. They
therefore form an Abelian group isomorphic to
$\mathbb{Z}_2^{\binom{n}{2}}$, providing a mechanism for
combining distributed graph inputs within the RGSP setup.
This composition property alone does not establish blindness
or security against colluding participants; those properties
depend on the surrounding protocol and its security
assumptions.

\subsection*{Example: adding a single edge}

Consider the three-qubit path with edges $(1,2)$ and $(2,3)$ (Fig.~\ref{fig:graph_transformation}),
whose adjacency matrix is:
\begin{equation}
A=
\begin{pmatrix}
0 & 1 & 0\\
1 & 0 & 1\\
0 & 1 & 0
\end{pmatrix}.
\end{equation}
To remove edge $(1,2)$, add edge $(1,3)$, and preserve edge
$(2,3)$, choose
\begin{equation}
\Delta=
\begin{pmatrix}
0 & 1 & 1\\
1 & 0 & 0\\
1 & 0 & 0
\end{pmatrix}.
\end{equation}
The corresponding phase update is:
\begin{equation}
\phi_{A'}(\vec{x})-\phi_A(\vec{x})
=
\pi(x_1x_2+x_1x_3)
\quad(\mathrm{mod}\;2\pi),
\end{equation}
and the transformed adjacency matrix is
\begin{equation}
A'=A\oplus\Delta
=
\begin{pmatrix}
0 & 0 & 1\\
0 & 0 & 1\\
1 & 1 & 0
\end{pmatrix}.
\end{equation}
The resulting edges are $(1,3)$ and $(2,3)$, confirming
the intended removal, addition, and preservation.

The same transformation can be distributed between two
clients: the first toggles only $(1,2)$ and the second
toggles only $(1,3)$. Their patterns satisfy
$\Delta^{(1)}\oplus\Delta^{(2)}=\Delta$, and each operation
can be implemented without knowledge of the other client's
pattern or the intermediate graph.

\section{Resource analysis for reduced-SWAP UBQC}
\label{app:resource}
We analyze the resource advantage enabled by RGSP-assisted blind quantum computation for the Quantum Fourier Transform (QFT) circuit~\cite{Barak2007}.

In standard BQC, the computation must be compiled onto a nearest-neighbor resource state (e.g., a 2D brickwork state~\cite{Briegel2001}). Because the brickwork state lacks long-range connectivity, non-local logical gates in the QFT require a dense arrangement of SWAP operations, which increases the circuit depth and the number of remotely prepared qubits.

Using the MasonQ compiler, the number of remotely prepared qubits required for a brickwork-state pattern is Eq.~\eqref{eq:brickwork_resource}. In contrast, RGSP directly prepares the target graph topology, including long-range connections. The target graph size is determined strictly by the logical depth of the QFT circuit, completely eliminating the SWAP overhead within the state preparation phase. The resource requirements for QFT sizes $n=3$ to $n=6$ are compared in Table~\ref{tab:resource_comparison_appendix}.

\begin{table*}
\centering
\begin{tabular}{c|c|c|c}
\hline
\textbf{QFT Size} & \textbf{RGSP (no SWAPs)} & \textbf{Standard RSP (with SWAPs)} & \textbf{Resource Advantage} \\ \hline
QFT3 & $6 \times 4 = 24$ & $26 \times 4 = 104$ & 76.9\% \\
QFT4 & $8 \times 4 = 32$ & $45 \times 4 = 180$ & 79.1\% \\
QFT5 & $10 \times 4 = 40$ & $70 \times 4 = 280$ & 81.4\% \\
QFT6 & $12 \times 4 = 48$ & $105 \times 4 = 420$ & \textbf{88.6\%} \\ \hline
\end{tabular}
\caption{Remotely prepared qubit resources for QFT implementations using RGSP and standard sequential RSP. RGSP directly prepares the target graph without SWAP operations, whereas standard RSP uses a fixed universal resource with SWAP-based routing. The resource advantage denotes the relative reduction in remotely prepared qubits, $1-N_{\mathrm{RGSP}}/N_{\mathrm{RSP}}$.}
\label{tab:resource_comparison_appendix}
\end{table*}

\begin{figure}
    \centering
    \hspace*{-2.0cm}
    \includegraphics[width=\columnwidth]{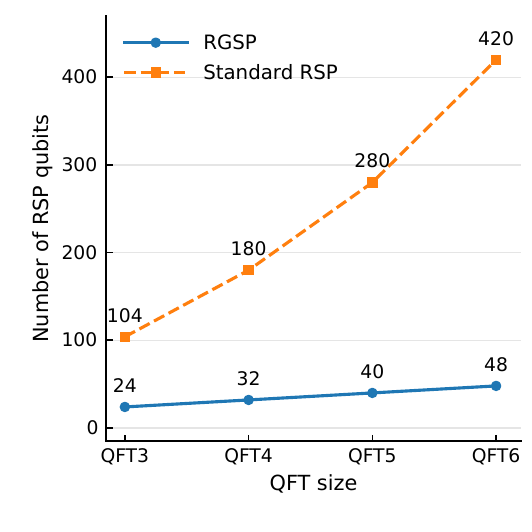}
    \caption{ Number of qubits required for QFT resource-state preparation. The curves correspond to the data in Table~\ref{tab:resource_comparison_appendix}. The blue curve shows RGSP, while the orange curve shows conventional RSP with SWAP overhead. The resource savings provided by RGSP become more significant as the QFT size increases.
}
\label{fig:qft_rsp_resource_comparison}
\end{figure}

As shown in Fig.~\ref{fig:qft_rsp_resource_comparison}, the resource advantage grows rapidly with system size, demonstrating that RGSP-assisted BQC offers an exceptionally scalable path for executing algorithms with high long-range connectivity.

\section{RGSP for Quantum Position Verification (QPV)}
\label{app:qpv}
Quantum Position Verification (QPV) exploits relativistic timing constraints and the no-cloning theorem to authenticate a prover's location~\cite{Kent2011a,Chandran2014, Brassard2011,Kent2011b}. Traditional QPV protocols (e.g., $\text{QPV}_{f\text{-BB84}}$) prepare and transmit qubits sequentially~\cite{Bluhm2022}. This sequential approach, however, faces a rate bottleneck due to channel losses in optical fibers. If a protocol requires $n$ logical qubits to be received and processed simultaneously, such as in parallel QPV~\cite{escolàfarràs2025quantumpositionverificationshot} or moving-prover scenarios like the Quantum Boat protocol~\cite{bartusek2026trackqubitsspacetime}, the success probability of the sequential qubit-by-qubit transmission scales as $\eta_t^n$, where $\eta_t$ is the single-photon transmission efficiency. This exponential suppression makes long-distance verification practically infeasible even for small $n$.

\begin{figure}
    \centering
    \hspace*{-2.0cm}
    \includegraphics[width=0.9\columnwidth]{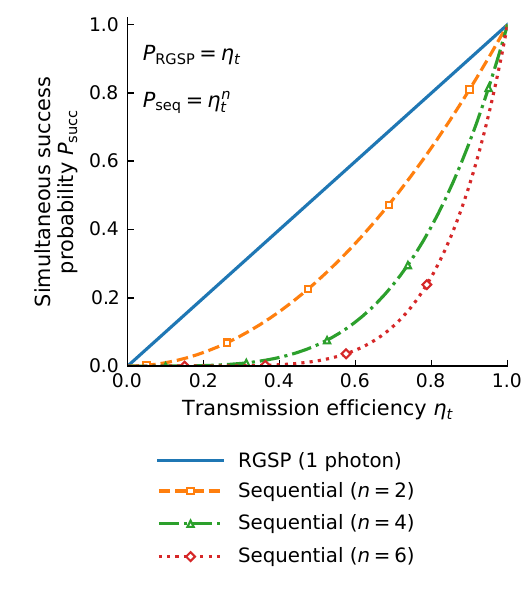}
    \caption{Success probability $P_{\mathrm{succ}}$ versus transmission efficiency $\eta_t$ for a fixed acceptance window. Sequential transmission of $n$ qubits scales as $P_{\mathrm{seq}}=\eta_t^n$, while RGSP, encoding all $n$ qubits in a single photon, scales as $P_{\mathrm{RGSP}}=\eta_t$ when internal routing and detection efficiencies are set to unity. Curves are shown for $n=2,4,6$.
    }
    \label{fig:success_eta}
\end{figure}

RGSP provides a window-constrained rate advantage by distributing the complete $n$-qubit logical state within a single high-dimensional photon. Because the $n$ logical qubits are encoded across $2^n$ temporal modes of a single qudit, the success probability of the protocol depends only on the transmission and detection of a single photon:
\begin{equation}
P_{\text{RGSP}} \propto \eta_t,
\label{eq:qpv_rgsp_scaling}
\end{equation}
which scales \emph{linear} with the transmission efficiency (Fig.~\ref{fig:success_eta}). This linear scaling bypasses the exponential loss bottleneck, increasing the multi-qubit verification rate by several orders of magnitude over long-distance fibers.

Second, we propose a QPV protocol based on graph-to-graph transformations introduced in Appendix \ref{app:g2g}. In this scheme, the verifier $V_0$ prepares a single-photon state with a linear phase function $\phi_x(\vec{\theta}_0) = \vec{\theta}_0 \cdot \vec{x}$, while $V_1$ sends a classical phase shift $\phi_{1,x} = \vec{\theta}_1 \cdot \vec{x}$. The prover $P$ at $pos^*$ is challenged to perform a non-local transformation $T_{\Delta}$ corresponding to a toggle matrix $\Delta = A \oplus A'$. By applying the quadratic phase shift $\pi \sum \Delta_{ij} x_i x_j$ to the propagating qudit in a single-shot, the prover implements a graph-to-graph transformation that updates the state to the target graph state:
\begin{equation}
|\Psi( \vec{\theta}_0 + \vec{\theta}_1 ) \rangle = \frac{1}{2^{n/2}} \sum_{\vec{x}} e^{i(\vec{\theta}_0 + \vec{\theta}_1) \cdot \vec{x} + i\pi \sum \Delta_{ij} x_i x_j} |1\rangle_x ,
\label{eq:qpv_g2g_state}
\end{equation}
and returns it immediately. The verifiers check the timing consistency and perform a projection measurement onto the target graph state. Because the transformation $T_{\Delta}$ belongs to an Abelian group, it cannot be performed by spatially separated adversaries (Alice and Bob) without violating relativistic timing constraints. Proving the ability to execute this multi-qubit graph-to-graph transformation in a single shot therefore provides robust, loss-tolerant position authentication.

\begin{figure*}[htbp]
    \centering
    \includegraphics[width=0.9\textwidth]{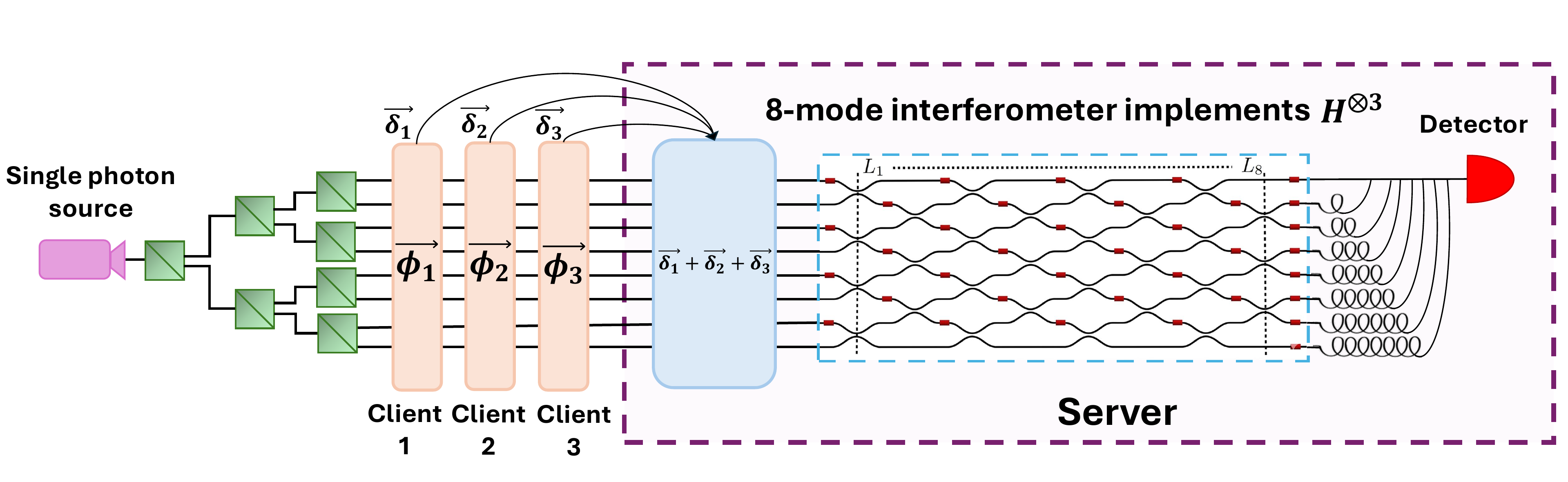}
    \caption{Using RGSP in a multi-party architecture in the style of QLine~\cite{polacchiMulticlientDistributedBlind2023}. A single photon is distributed across \(2^n\) optical modes and sequentially acquires secret phase contributions from multiple clients. The accumulated phase pattern encodes both the local computational angles and the graph-state correlations. For a near term implementation in the absence of a quantum server that holds a memory register, a \(2^n\)-mode interferometer implementing \(H^{\otimes n}\) mixes all optical modes coherently, and the detector samples the output probability distribution corresponding to the delegated computation.}
    \label{fig:qline_grsp_architecture}
\end{figure*} 

\section{Multi-party delegated computation}
\label{multi_client}
This appendix shows how the computational layer of the QLine architecture can be reformulated within the phase-function framework developed in Fig.~\ref{fig:qline_grsp_architecture}.  A single photon is initially prepared in an equal superposition of $2^n$ temporal modes. As the photon propagates sequentially along the QLine through $N$ clients, each client $k$ applies its secret local phase rotation $\alpha_i^{(k)}$ and one-time-pad bit $z_i^{(k)}$ to the corresponding temporal modes. The accumulated phase pattern becomes:
\begin{equation}
\Phi_x = \sum_{i=1}^n \Theta_i x_i + \pi \sum_{i<j, k} A_{ij}^{(k)} x_i x_j \pmod{2\pi}
\label{eq:qline_phase_state}
\end{equation}
where $\Theta_i = \sum_k (\alpha_i^{(k)} + \pi z_i^{(k)})$ represents the joint computational angles. 

This accumulated phase function has the exact same binary quadratic structure as the RGSP phase-encoding rule [Eq.~(\ref{eq:phase-rule})], meaning that both the local computational rotations and the collaborative graph entanglement are encoded \emph{directly} into the photon's phase profile. At the end of the line, the server performs a $2^n$-mode interferometer (implementing a multi-dimensional Hadamard transform $H^{\otimes n}$) to coherently mix all temporal modes. The detector click statistics then sample the encrypted probability distribution $P(\vec{y})$ corresponding to the delegated multi-party computation. 

By classically combining their private OTP keys and correction parameters, the clients can jointly decode the server's measurement statistics to recover the final outcome, while keeping the computation perfectly blind to the server. Because the entanglement is entirely intra-particle and no multi-photon coincidence is required, this framework allows secure multi-party quantum computation for \emph{more than two clients} on near-term NISQ networks without the overhead of QEC or the losses of multi-photon states.

\onecolumngrid

\subsection*{Example: Two-Qubit Case}
\label{sec:two_qubit_example}

We first consider a two-qubit fragment of the QLine computation. The corresponding circuit can be written as:
\begin{equation}
(M_X\otimes M_X)
\bigl(R_z(\alpha_1)\otimes R_z(\alpha_2)\bigr)
CZ_{12}
\bigl(Z^{z_1}\otimes Z^{z_2}\bigr)
|+\rangle|+\rangle.
\end{equation}

The initial two-qubit state is:
\begin{equation}
|+\rangle|+\rangle = \frac12 \sum_{x_1,x_2\in{0,1}} |x_1x_2\rangle.
\end{equation}

\paragraph*{Encoding the Classical Bit-Flip Data}

We first encode the classical bit-flip data through the local $Z$ operations. Applying $(Z^{z_1}\otimes Z^{z_2})$, we obtain:
\begin{equation}
\begin{aligned}
(Z^{z_1}\otimes Z^{z_2})|x_1x_2\rangle
&= e^{i\pi z_1x_1} e^{i\pi z_2x_2} |x_1x_2\rangle= e^{i\pi(z_1x_1+z_2x_2)} |x_1x_2\rangle.
\end{aligned}
\end{equation}

Thus,
\begin{equation}
|\Psi_1\rangle = \frac12 \sum_{x_1,x_2} e^{i\pi(z_1x_1+z_2x_2)} |x_1x_2\rangle.
\end{equation}

\begin{figure}[t]
\begin{tcolorbox}[
    width=\columnwidth,
    colback=white,
    colframe=black,
    boxrule=0.5pt,
    arc=0pt,
    left=7pt,
    right=7pt,
    top=5pt,
    bottom=5pt
]

\textbf{Multi-party blind quantum computation}

\vspace{0.08cm}
\raggedright
\textbf{Input:}

\begin{itemize}
    \setlength{\itemsep}{1pt}
    \setlength{\parskip}{0pt}
    \item Client $k$ has classical data $z_i^{(k)}$ and effective angles $\alpha_i^{(k)}$.
    \item The $n$ clients collectively specify a graph $G=(V,E)$.
    \item The server has no private input.
\end{itemize}

\textbf{Protocol:}

\begin{enumerate}
    \setlength{\itemsep}{1pt}
    \setlength{\parskip}{0pt}
    \item A single-photon source creates an equal superposition over $2^n$ optical modes.
    \item The $n$ clients sequentially apply their inputs.
    \item After all clients have acted, the phase-encoded state $|\Psi\rangle$ is sent to the server.
    \item The server applies a $2^n$-mode interferometer implementing $H^{\otimes n}$.
    \item The interferometer produces output amplitudes $A(\vec{y})$ and probabilities $P(\vec{y})$.
    \item The detector click statistics reproduce the output distribution of the original QLine computation.
\end{enumerate}

\end{tcolorbox}
\end{figure}

\paragraph*{Encoding the Rotation Angles}

Next, we encode the local rotation angles. Using the phase convention $(R_z(\alpha)|x\rangle = e^{i\alpha x}|x\rangle)$, we have:
\begin{equation}
\begin{aligned}
\bigl(R_z(\alpha_1)\otimes R_z(\alpha_2)\bigr) |x_1x_2\rangle
&= e^{i\alpha_1 x_1} e^{i\alpha_2 x_2} |x_1x_2\rangle
= e^{i(\alpha_1 x_1+\alpha_2 x_2)} |x_1x_2\rangle.
\end{aligned}
\end{equation}

Therefore,
\begin{equation}
|\Psi_2\rangle = \frac12 \sum_{x_1,x_2} e^{i[\pi(z_1x_1+z_2x_2) + \alpha_1 x_1+\alpha_2 x_2]} |x_1x_2\rangle.
\end{equation}

\paragraph*{Encoding the Graph Connectivity}

For two qubits, the graph contains one edge $(1,2)$. Applying $CZ_{12}$, we obtain:
\begin{equation}
CZ_{12}|x_1x_2\rangle = e^{i\pi x_1x_2}|x_1x_2\rangle.
\end{equation}

The resulting phase-encoded state is:
\begin{equation}
|\Psi\rangle = \frac12 \sum_{x_1,x_2\in{0,1}} e^{i\Phi(x_1,x_2)} |x_1x_2\rangle,
\end{equation}
with
\begin{equation}
\Phi(x_1,x_2) = (\pi z_1+\alpha_1)x_1 + (\pi z_2+\alpha_2)x_2 + \pi x_1x_2.
\end{equation}

The phase function contains three distinct contributions:
\begin{itemize}
\item $\pi z_i x_i$ from the classical bit-flip $Z^{z_i}$,
\item $\alpha_i x_i$ from the local rotation $R_z(\alpha_i)$,
\item $\pi x_1x_2$ from the graph entanglement $CZ_{12}$.
\end{itemize}

This decomposition makes explicit that the local computational parameters and the graph connectivity enter the state through additive phase contributions.

\paragraph*{Explicit Expansion}

Evaluating $(\Phi(x_1,x_2))$ on all computational-basis strings gives:
\begin{equation}
\begin{aligned}
\Phi(00) &= 0,\quad 
\Phi(01) = \pi z_2+\alpha_2, \quad
\Phi(10) = \pi z_1+\alpha_1, \quad
\Phi(11) = \pi(z_1+z_2)+\alpha_1+\alpha_2+\pi.
\end{aligned}
\end{equation}

Hence,
\begin{equation}
\begin{aligned}
|\Psi\rangle = \frac12\big(
&|00\rangle + e^{i(\pi z_2+\alpha_2)}|01\rangle + e^{i(\pi z_1+\alpha_1)}|10\rangle
+ e^{i[\pi(z_1+z_2)+\alpha_1+\alpha_2+\pi]}|11\rangle \big).
\end{aligned}
\end{equation}

\paragraph*{Two-Qubit State as a Four-Mode Qudit}

The two-qubit computational basis can be mapped to a four-dimensional photonic qudit according to:
\begin{equation}
|00\rangle \leftrightarrow |0\rangle,\quad
|01\rangle \leftrightarrow |1\rangle,\quad
|10\rangle \leftrightarrow |2\rangle,\quad
|11\rangle \leftrightarrow |3\rangle,
\end{equation}
so that the state becomes a single-photon state distributed over four interferometer modes:
\begin{equation}
|\Psi\rangle = \frac12 \sum_{\vec{x}\in{0,1}^2} e^{i\Phi(x)} |\vec{x}\rangle_{\mathrm{qudit}}.
\end{equation}

Thus, the classical QLine data, local rotation angles, and graph connectivity are encoded in the phases of the interferometer modes.

\paragraph*{Two-Qubit Output Statistics}

The measurement of both qubits in the $X$-basis is equivalent to applying $H\otimes H$ followed by computational-basis measurement. For one qubit,
\begin{equation}
H|x_i\rangle = \frac{1}{\sqrt{2}} \sum_{y_i\in{0,1}} (-1)^{y_i x_i}|y_i\rangle.
\end{equation}

Therefore,
\begin{equation}
(H\otimes H)|x_1x_2\rangle = \frac12 \sum_{y_1,y_2} (-1)^{y_1 x_1 + y_2 x_2} |y_1y_2\rangle.
\end{equation}

Applying this transformation to $|\Psi\rangle$, we obtain:
\begin{align}
(H\otimes H)|\Psi\rangle
&= \frac12 \sum_{x_1,x_2} e^{i\Phi(x_1,x_2)} (H\otimes H)|x_1x_2\rangle
= \frac14 \sum_{x_1,x_2} \sum_{y_1,y_2} (-1)^{y_1 x_1 + y_2 x_2} e^{i\Phi(x_1,x_2)} |y_1y_2\rangle.
\end{align}

The output amplitude for $y=(y_1,y_2)$ is:
\begin{equation}
A(y_1,y_2) = \frac14 \sum_{x_1,x_2\in{0,1}} (-1)^{y_1 x_1 + y_2 x_2} e^{i\Phi(x_1,x_2)}.
\end{equation}
with probability $(P(y_1,y_2)=|A(y_1,y_2)|^2)$.

Writing $\Phi_{ab} \equiv \Phi(a,b)$, the four amplitudes are:

\begin{equation}
\begin{aligned}
A(00) &= \tfrac14 \big( e^{i\Phi_{00}} + e^{i\Phi_{01}} + e^{i\Phi_{10}} + e^{i\Phi_{11}} \big), \\
A(01) &= \tfrac14 \big( e^{i\Phi_{00}} - e^{i\Phi_{01}} + e^{i\Phi_{10}} - e^{i\Phi_{11}} \big), \\
A(10) &= \tfrac14 \big( e^{i\Phi_{00}} + e^{i\Phi_{01}} - e^{i\Phi_{10}} - e^{i\Phi_{11}} \big), \\
A(11) &= \tfrac14 \big( e^{i\Phi_{00}} - e^{i\Phi_{01}} - e^{i\Phi_{10}} + e^{i\Phi_{11}} \big).
\end{aligned}
\end{equation}

Each output amplitude is therefore determined by the coherent interference of the four phase-encoded input modes, and the corresponding output probability is obtained from its squared modulus.

\subsection*{Example: Three-Qubit Case}
\label{sec:three_qubit_example}

As an explicit three-qubit example, consider a linear graph with edges $E={(1,2),(2,3)}$. The corresponding QLine computation can be represented by an eight-mode photonic interferometer with mode-dependent phase shifts.

The initial state is:
\begin{equation}
|+\rangle^{\otimes 3} = \frac{1}{\sqrt{8}} \sum_{x_1,x_2,x_3\in{0,1}} |x_1x_2x_3\rangle.
\end{equation}

\paragraph*{Encoding the Classical Bit-Flip Data}

Applying $Z^{z_1}\otimes Z^{z_2}\otimes Z^{z_3}$, we obtain:
\begin{align}
(Z^{z_1}\otimes Z^{z_2}\otimes Z^{z_3}) |x_1x_2x_3\rangle
&= e^{i\pi(z_1 x_1+z_2 x_2+z_3 x_3)} |x_1x_2x_3\rangle.
\end{align}

Thus,
\begin{equation}
|\Psi_1\rangle = \frac{1}{\sqrt{8}} \sum_{x\in{0,1}^3} e^{i\pi(z_1 x_1+z_2 x_2+z_3 x_3)} |x_1x_2x_3\rangle.
\end{equation}

\paragraph*{Encoding the Rotation Angles}

We next encode the local rotation angles. Applying $R_z(\alpha_1)\otimes R_z(\alpha_2)\otimes R_z(\alpha_3)$, we have:
\begin{align}
\bigl(R_z(\alpha_1)\otimes R_z(\alpha_2)\otimes R_z(\alpha_3)\bigr) |x_1x_2x_3\rangle
= e^{i(\alpha_1 x_1+\alpha_2 x_2+\alpha_3 x_3)} |x_1x_2x_3\rangle.
\end{align}

Therefore,
\begin{equation}
|\Psi_2\rangle = \frac{1}{\sqrt{8}} \sum_{\vec{x}\in\{0,1\}^3} e^{i[\pi(z_1 x_1+z_2 x_2+z_3 x_3) + \alpha_1 x_1+\alpha_2 x_2+\alpha_3 x_3]} |x_1x_2x_3\rangle.
\end{equation}

\paragraph*{Encoding the Graph Connectivity}

For the linear graph $1-2-3$ with edges $E={(1,2),(2,3)}$, the graph-entangling operation is $CZ_{12}CZ_{23}$:
\begin{equation}
CZ_{12} \Rightarrow e^{i\pi x_1 x_2}, \qquad CZ_{23} \Rightarrow e^{i\pi x_2 x_3}.
\end{equation}

The resulting three-qubit phase state is:
\begin{equation}
|\Psi\rangle = \frac{1}{\sqrt{8}} \sum_{\vec{x}\in\{0,1\}^3} e^{i\Phi(x)} |x_1x_2x_3\rangle,
\end{equation}
with
\begin{equation}
\Phi(x_1,x_2,x_3) = \sum_{i=1}^{3}(\pi z_i+\alpha_i)x_i + \pi x_1 x_2 + \pi x_2 x_3.
\end{equation}

Equivalently,
\begin{equation}
\Phi(x_1,x_2,x_3) = (\pi z_1+\alpha_1)x_1 + (\pi z_2+\alpha_2)x_2 + (\pi z_3+\alpha_3)x_3 + \pi x_1 x_2 + \pi x_2 x_3.
\end{equation}

The phase function therefore contains the three local contributions and the two quadratic terms associated with the edges of the linear graph.

\paragraph*{Explicit Expansion}

Evaluating $\Phi$ on all eight basis strings yields:
\begin{equation}
\begin{aligned}
\Phi(000) &= 0, \\
\Phi(001) &= \pi z_3+\alpha_3, \\
\Phi(010) &= \pi z_2+\alpha_2, \\
\Phi(011) &= \pi(z_2+z_3)+\alpha_2+\alpha_3+\pi, \\
\Phi(100) &= \pi z_1+\alpha_1, \\
\Phi(101) &= \pi(z_1+z_3)+\alpha_1+\alpha_3, \\
\Phi(110) &= \pi(z_1+z_2)+\alpha_1+\alpha_2+\pi, \\
\Phi(111) &= \pi(z_1+z_2+z_3)+\alpha_1+\alpha_2+\alpha_3+2\pi.
\end{aligned}
\end{equation}

Since $e^{i2\pi}=1$, we have $\Phi(111) \equiv \pi(z_1+z_2+z_3)+\alpha_1+\alpha_2+\alpha_3 \pmod{2\pi}$.

Therefore,
\begin{equation}
\begin{aligned}
|\Psi\rangle = \frac{1}{\sqrt{8}}\big(
&|000\rangle + e^{i(\pi z_3+\alpha_3)}|001\rangle + e^{i(\pi z_2+\alpha_2)}|010\rangle \\
&+ e^{i[\pi(z_2+z_3)+\alpha_2+\alpha_3+\pi]}|011\rangle + e^{i(\pi z_1+\alpha_1)}|100\rangle \\
&+ e^{i[\pi(z_1+z_3)+\alpha_1+\alpha_3]}|101\rangle + e^{i[\pi(z_1+z_2)+\alpha_1+\alpha_2+\pi]}|110\rangle \\
&+ e^{i[\pi(z_1+z_2+z_3)+\alpha_1+\alpha_2+\alpha_3]}|111\rangle \big).
\end{aligned}
\end{equation}

\paragraph*{Three-Qubit State as an Eight-Mode Qudit}

The three-qubit computational basis can be mapped to an eight-dimensional qudit according to:
\begin{align}
|000\rangle &\leftrightarrow |0\rangle,\quad |001\rangle \leftrightarrow |1\rangle,\quad |010\rangle \leftrightarrow |2\rangle,\quad |011\rangle \leftrightarrow |3\rangle,\
|100\rangle &\leftrightarrow |4\rangle,\quad |101\rangle \leftrightarrow |5\rangle,\quad |110\rangle \leftrightarrow |6\rangle,\quad |111\rangle \leftrightarrow |7\rangle.
\end{align}

Thus,
\begin{equation}
|\Psi\rangle = \frac{1}{\sqrt{8}} \sum_{\vec{x}\in\{0,1\}^3} e^{i\Phi(x)} |\vec{x}\rangle_{\mathrm{qudit}}.
\end{equation}

This representation corresponds to a single photon distributed over eight optical modes, with each mode carrying a phase determined by the QLine classical data, the local rotation angles, and the graph connectivity.

\paragraph{Three-Qubit Output Statistics}

Measurement of all three qubits in the (X)-basis is equivalent to applying $H^{\otimes 3}$ followed by computational-basis measurement. For a single qubit,
\begin{equation}
H|x_i\rangle = \frac{1}{\sqrt{2}} \sum_{y_i\in\{0,1\}} (-1)^{y_i x_i}|y_i\rangle.
\end{equation}

Therefore,
\begin{equation}
H^{\otimes 3}|x_1x_2x_3\rangle = \frac{1}{\sqrt{8}} \sum_{y_1,y_2,y_3} (-1)^{y_1 x_1 + y_2 x_2 + y_3 x_3} |y_1y_2y_3\rangle.
\end{equation}

With binary inner product $y\cdot x = y_1 x_1 + y_2 x_2 + y_3 x_3$, we have:
\begin{equation}
H^{\otimes 3}|\vec{x}\rangle = \frac{1}{\sqrt{8}} \sum_{\vec{y}\in\{0,1\}^3} (-1)^{y\cdot x}|\vec{y}\rangle.
\end{equation}

Applying this transformation to $|\Psi\rangle$, we obtain:
\begin{align}
H^{\otimes 3}|\Psi\rangle
&= \frac{1}{\sqrt{8}} \sum_{\vec{x}} e^{i\Phi(x)} H^{\otimes 3}|\vec{x}\rangle, \
&= \frac{1}{\sqrt{8}} \sum_{\vec{x}} e^{i\Phi(x)} \left[ \frac{1}{\sqrt{8}} \sum_{\vec{y}} (-1)^{y\cdot x}|\vec{y}\rangle \right], \
&= \frac{1}{8} \sum_{\vec{y}} \left( \sum_{\vec{x}} (-1)^{y\cdot x} e^{i\Phi(x)} \right) |\vec{y}\rangle.
\end{align}

Hence the output amplitude for $\vec{y}=(y_1,y_2,y_3)$ is:
\begin{equation}
A(y) = \frac{1}{8} \sum_{\vec{x}\in\{0,1\}^3} (-1)^{y\cdot x} e^{i\Phi(x)}.
\end{equation}
with probability $P(y)=|A(y)|^2$.

\paragraph*{Explicit Amplitudes}

Writing $\Phi_{abc} \equiv \Phi(a,b,c)$, the eight output amplitudes are:
\begin{equation}
\begin{aligned}
A(000) &= \tfrac18 \big( e^{i\Phi_{000}} + e^{i\Phi_{001}} + e^{i\Phi_{010}} + e^{i\Phi_{011}} + e^{i\Phi_{100}} + e^{i\Phi_{101}} + e^{i\Phi_{110}} + e^{i\Phi_{111}} \big), \\
A(001) &= \tfrac18 \big( e^{i\Phi_{000}} - e^{i\Phi_{001}} + e^{i\Phi_{010}} - e^{i\Phi_{011}} + e^{i\Phi_{100}} - e^{i\Phi_{101}} + e^{i\Phi_{110}} - e^{i\Phi_{111}} \big), \\
A(010) &= \tfrac18 \big( e^{i\Phi_{000}} + e^{i\Phi_{001}} - e^{i\Phi_{010}} - e^{i\Phi_{011}} + e^{i\Phi_{100}} + e^{i\Phi_{101}} - e^{i\Phi_{110}} - e^{i\Phi_{111}} \big), \\
A(011) &= \tfrac18 \big( e^{i\Phi_{000}} - e^{i\Phi_{001}} - e^{i\Phi_{010}} + e^{i\Phi_{011}} + e^{i\Phi_{100}} - e^{i\Phi_{101}} - e^{i\Phi_{110}} + e^{i\Phi_{111}} \big), \\
A(100) &= \tfrac18 \big( e^{i\Phi_{000}} + e^{i\Phi_{001}} + e^{i\Phi_{010}} + e^{i\Phi_{011}} - e^{i\Phi_{100}} - e^{i\Phi_{101}} - e^{i\Phi_{110}} - e^{i\Phi_{111}} \big), \\
A(101) &= \tfrac18 \big( e^{i\Phi_{000}} - e^{i\Phi_{001}} + e^{i\Phi_{010}} - e^{i\Phi_{011}} - e^{i\Phi_{100}} + e^{i\Phi_{101}} - e^{i\Phi_{110}} + e^{i\Phi_{111}} \big), \\
A(110) &= \tfrac18 \big( e^{i\Phi_{000}} + e^{i\Phi_{001}} - e^{i\Phi_{010}} - e^{i\Phi_{011}} - e^{i\Phi_{100}} - e^{i\Phi_{101}} + e^{i\Phi_{110}} + e^{i\Phi_{111}} \big), \\
A(111) &= \tfrac18 \big( e^{i\Phi_{000}} - e^{i\Phi_{001}} - e^{i\Phi_{010}} + e^{i\Phi_{011}} - e^{i\Phi_{100}} + e^{i\Phi_{101}} + e^{i\Phi_{110}} - e^{i\Phi_{111}} \big).
\end{aligned}
\end{equation}

Each output probability is $P(abc)=|A(abc)|^2$.

The two- and three-qubit examples make the correspondence between the original QLine description~\cite{polacchiMulticlientDistributedBlind2023} and the phase-encoded photonic representation explicit. In both cases, the local $Z$-type classical data, single-qubit rotations, and graph connectivity contribute additively to the phase function. The resulting phase profile is then transformed by the $H^{\otimes n}$ interferometer, with the detector output determined by the corresponding interference amplitudes.

In the original QLine protocol, the output of the delegated computation is obtained from measurement outcomes on a graph state. These outcomes are protected by the blindness mechanisms of the protocol, including random OTP rotations and classical correction bits contributed by the clients.

The same computational structure is retained in the interferometric formulation. The phase function $\Phi(x)$ contains both the computational parameters and the blindness-related variables. Consequently, the server accesses only the detector statistics associated with the phase-encoded output distribution $P(\vec{y})$. 

After the protocol is completed, the clients use their private classical information, including the random padding variables and correction data, to interpret the observed outcomes and recover the intended computational output. Under the blindness assumptions of the original QLine protocol, the server does not obtain the private information required to reconstruct the underlying computation from this data alone.

Thus, the interferometric formulation preserves the computational structure of the original QLine protocol while expressing the multi-client inputs and graph-state correlations through a common phase-function representation.
\end{document}